\documentclass[12pt,preprint]{aastex}

 \def\Jbold{{\bf J}} 
\def\Hcal{{\cal H}}

\def\apgt{\ {\raise-.5ex\hbox{$\buildrel>\over\sim$}}\ }
\def\aplt{\ {\raise-.5ex\hbox{$\buildrel<\over\sim$}}\ }

\def\Hcal{{\cal H}}  \def\Fcal{{\cal F}}
 \def\Acal{{\cal A}}

\def\Hac{\Hcal_{\rm ac}}

\def\Rg{{\cal R}_g}

\def\ex{\rm e}

\def\apgt{\ {\raise-.5ex\hbox{$\buildrel>\over\sim$}}\ }
\def\aplt{\ {\raise-.5ex\hbox{$\buildrel<\over\sim$}}\ }

\shorttitle{Evolution and impact of Comet 9P/Tempel 1}
\shortauthors{Sarid, Prialnik, Meech, Pittichov\'{a}, Farnham}

\slugcomment{Submitted April 12, 2005 ; Accepted May 6, 2005 (to PASP)}

\begin{document}

\title{Thermal evolution and activity of Comet 9P/Tempel 1 \\ and simulation of a deep impact}

\author{
Gal Sarid and Dina Prialnik,\altaffilmark{1} \\
Karen J. Meech and Jana Pittichov\'{a},\altaffilmark{2}\\
and\\
Tony L. Farnham\altaffilmark{3}
}

\altaffiltext{1}{Department of Geophysics and Planetary Sciences,
Sackler Faculty of Exact Sciences, Tel Aviv University, Ramat Aviv
69978, Israel}
\altaffiltext{2}{Institute for Astronomy, 2680 Woodlawn Drive, Honolulu, HI  96822}
\altaffiltext{3}{Department of Astronomy, University of Maryland, College Park, MD 20742-2421}

\begin{abstract}

   We use a quasi 3-D thermal evolution model for a spherical comet
   nucleus, which takes into account the diurnal and latitudinal variation
   of the solar flux, but neglects lateral heat conduction. We model the
   thermal evolution and activity of Comet 9P/Tempel 1, in anticipation
   of the {\it Deep Impact} mission encounter with the comet. We also investigate the
   possible outcome of a projectile impact, assuming that all the energy is absorbed
   as thermal energy. An interesting result of this investigation, is that the estimated
   amount of dust ejected due to the impact is equivalent to 2--2.6 days of activity, during
   "quiet" conditions, at perihelion.

   We show that production rates of volatiles that are released in the interior
   of the nucleus depend strongly on the porous structure, in particular on
   the surface to volume ratio of the pores. We develop a more accurate model
   for calculating this parameter, based on a distribution of pore sizes, rather
   than a single, average pore size.

\end{abstract}


\section{Introduction}
\label{introduction}

\noindent
Comet 9P/Tempel 1 is the target of NASA's {\it Deep Impact} mission (Belton and A'Hearn 1999,
Meech et al. 2000, A'Hearn et al. 2005). The spacecraft, successfully launched on January 12, 2005,
and due to encounter the comet on UT 5:52 ($\pm$ 3 min) July 4, 2005, will collect images of
the nucleus, and release a smaller projectile spacecraft.
The projectile will maneuver its path for a collision with the nucleus of the target comet.

Comet 9P/Tempel 1 was first discovered in 1867, and
recovered in 1873 and 1879. In 1881, the comet passed close to Jupiter, its orbit changed and
it was lost. Later investigations (Marsden 1963) revealed that it had undergone two more
close encounters with Jupiter, in 1941 and 1957. The comet reappeared in 1967 and since that time
it has been observed at every apparition. Its present orbital period is 5.5 years and the
perihelion distance is 1.506 AU.

In anticipation of the mission, comet 9P/Tempel 1 has been intensively observed by both professional
and amateur astronomers (Lamy et al. 2001, Meech 2002) and data about its size, rotation, production rates, etc.,
has accumulated (Fern\'{a}ndez et al. 2003, Belton et al. 2005).
On the theoretical side, models have been developed to simulate the impact and its consequences
(Schultz and Anderson 2005, W\"{u}nnemann et al. 2005).
It is impossible, however, to simulate the impact accurately,
since very little is known about the properties of cometary material. Clearly, the impact energy
--- originally kinetic --- will divide between mechanical and thermal energies, but the proportion
will be strongly determined by the nature of cometary material: its strength, porosity, thermal
conductivity, and so forth.

Since the dynamical side of the impact has already been investigated in detail (Nolan et al. 1996,
Housen 2002, Schultz et al. 2002), we focus in this paper on the thermal aspect.
We first choose an initial model that matches as well as possible the observations of comet 9P/Tempel~1.
This is achieved by running a number of models, for different assumptions and parameter combinations,
through several orbital revolutions. The composition of these models will be assumed to
include water ice,
dust and 5 additional volatile species. A "thermal" impact on this model will then be simulated.

The evolution --- both
long-term, prior and following the impact, and short-term during the event itself --- will be
calculated by means of a quasi-3D code (Prialnik et al. 2005). This code takes into account
diurnal and latitudinal variations, but neglects lateral heat conduction, and is an expanded
version of the code used by Cohen et al. (2003).

In Section \ref{code} we briefly describe the numerical code. Section \ref{newsvr} describes
improvements implemented to the code that take into account a pore size distribution, rather
than an average (fixed) pore size.
Next, in Section \ref{observations} we describe ground-based observations that
have been recently carried out in
anticipation of the {\it Deep Impact} mission in order to determine
the comet's activity levels as a function of heliocentric distance. We include a subset of the
observations in this paper for the purpose of modeling the dust coma so that we can obtain dust
fluxes and the heliocentric distances between which the comet is
active as constraints and tests for the thermal models.
In the following Section \ref{results} we show the results of
numerical calculations for the thermal evolution of comet 9P/Tempel~1 along several orbital revolutions.
The simulation of the impact and its consequences are described and discussed in Section \ref{simpact}.
Our main conclusions may be found in Section \ref{conclusions}.

\section{Numerical model and parameters}
\label{code}

\noindent
Our model assumes a porous, spherical and initially homogeneous nucleus
composed of amorphous and crystalline water ice, dust, and 5 other volatile species:
CO, CO$_2$, HCN, NH$_3$, and C$_2$H$_2$. The equations of energy and mass conservation
for this system are briefly summarized below (using standard notation, e.g., Prialnik et al. 2005).
\par
We use a quasi-3D approach, where diurnal and latitudinal temperature variations are calculated
as they result from uneven surface heating by solar radiation onto a spinning comet. Lateral
heat conduction is neglected, and so heat is assumed to flow radially (perpendicular to the surface).
Thus, different points on the surface do not interact. This simple approach is justified by
the low thermal conductivity of cometary ice mixtures and by the thinness of the skin-depth
(see Cohen et al. 2003).

The bulk mass density $\rho$ and porosity $\Psi$ are given, respectively, by
\begin{equation}
\rho=\rho_a+\rho_c+\rho_v+\sum_\alpha(\rho_{s,\alpha}+\rho_{g,\alpha})+\rho_d,
\end{equation}
\begin{equation}
\Psi=1-(\rho_a+\rho_c)/\varrho_{\rm ice}-\sum_\alpha\rho_{s,\alpha}/\varrho_\alpha
-\rho_d/\varrho_d,
\end{equation}
where $\varrho$ is the specific density of a species $\alpha$.
The meaning of indices is: $a$ - amorphous water ice, $c$ - crystalline water ice, $ice$ - when
the phases are indistinguishable, $v$ - water vapor,
$s$ and $g$ - solid and gaseous phases of a volatile species, respectively, and $d$ - dust.
The partial pressure, assuming an ideal gas, is
\begin{equation}
P_\alpha={\Rg\rho_{g,\alpha}T\over\Psi\mu}.
\end{equation}
Because amorphous ice has a tendency to convert to the crystalline form spontaneously, we have
\begin{equation}
{\partial\rho_a\over\partial t}=-\lambda\rho_a,
\label{droadt}
\end{equation}
where $\lambda$ is the rate of crystallization. If $f$ is the total fraction of occluded gas,
$\sum_\alpha f_\alpha=f$, the equations of mass conservation are
\begin{equation}
{\partial\rho_c\over\partial t}=(1-f)\lambda\rho_a-q_v,
\end{equation}
\begin{equation}
{\partial\rho_v\over\partial t}+\nabla\cdot {\bf J}_v=q_v,
\label{mcons}
\end{equation}
for H$_2$O in both phases, and similarly,
\begin{equation}
{\partial\rho_{g,\alpha}\over\partial t}+\nabla\cdot {\bf J}_\alpha=
f_\alpha\lambda\rho_a+q_\alpha, \qquad {\partial\rho_{s,\alpha}\over\partial t}=-q_\alpha,
\label{mconsgs}
\end{equation}
for the other volatile species.
Taking into account energy conservation throughout the nucleus, and combining with
the mass conservation equations, we obtain the heat diffusion equation in the form
\begin{equation}
\sum_\alpha\rho_\alpha{\partial u_\alpha\over\partial t}-\nabla\cdot(K\nabla T)
+\left(\sum_\alpha c_\alpha \Jbold_\alpha\right)\cdot\nabla T =
\lambda\rho_a\Hac-\sum_{\alpha}q_{\alpha}\Hcal_{\alpha},
\label{heateq}
\end{equation}
where $\Hac$ is the heat released upon crystallization, and $\Hcal_{\alpha}$ is the
heat of sublimation.
The above set of time-dependent equations
is subject to constitutive relations: $u(T)$, $\lambda(T)$,
$q_{\alpha}(T,\Psi,r_p)$, $\Jbold_{\alpha}(T,\Psi,r_p)$, $K(T,\Psi,r_p)$,
where $r_p$ denotes pore radius. These relations require additional assumptions for modeling
the structure of the nucleus.

Measurements by Schmitt et al. (1989) have shown that the rate of crystallization
is given by
\begin{equation}
\lambda (T)=1.05\times 10^{13}\ex^{-5370/T} \quad {\rm s}^{-1}.
\end{equation}
The rate of sublimation --- mass per unit volume of cometary material per unit time --- is given by
\begin{equation}
q_{\alpha}=S(\Psi,r_p)\left[(P_{{\rm vap},\alpha}(T)-P_{\alpha})\sqrt{\mu_{\alpha}\over2\pi R_gT}\ \right],
\label{subrate}
\end{equation}
where the term in square brackets represents the sublimation rate per unit surface area,
and $S$ represents the surface to volume ratio, which is a function of the given porosity and pore radius.
The saturated vapor pressure, $P_{\rm vap}$, is given by the
Clausius - Clapeyron equation:
\begin{equation}
P_{\rm vap}=P_0\exp^{-B/T},
\end{equation}
where $P_0=3.56\times10^{12}$~N~m$^{-2}$, and $B=6141.667$~K (Fanale and Salvail 1984).

Dust is assumed to be, in part, dragged along with the gas flowing through pores and, in part,
lifted off the nucleus surface by the sublimating vapor. For the former, the dust velocity is
assumed to be equal to the gas velocity (see Podolak and Prialnik 1996). An efficiency factor is
calculated, to take account of a dust size distribution that allows only grains up to a critical size to
be dragged or lifted off. The rest may accumulate to form a dust mantle. The efficiency factor may
be adjusted so as to allow or prevent the formation of a sealing mantle (see also Prialnik et al. 2005
and references therein).

The equations of mass conservation and energy transport are second-order in space and hence each require
two boundary conditions.
One boundary condition for eq.(\ref{heateq}) is vanishing heat flux at the center.
The second one is obtained from the requirement of energy balance at the surface:
\begin{equation}
F(R)=\epsilon\sigma T(R,t)^4\,+\,\Fcal P_{\rm vap}(T)\sqrt{\mu\over2\pi R_gT}\Hcal\,
-(1-\Acal){L_{\odot}\over 4\pi d_H(t)^2}\cos z.
\label{fluxs}
\end{equation}
In the simple case considered here, of a rotational axis that is perpendicular to the orbital plane,
the solar zenith angle is given in terms of latitude $\theta$, and hour
angle $\varphi=(2\pi/P_{\rm rot})t$,
\begin{equation}
\cos z=cos\theta\cos\varphi.
\end{equation}
The factor $\Fcal\le1$ represents the fractional area of exposed ice, since the surface material is
a mixture of ice and dust, and can be written as
\begin{equation}
\Fcal=\left(1+{\varrho_{\rm ice}\over\rho_{\rm ice}}{\rho_d\over\varrho_d}\right)^{-1}
\end{equation}
(Crifo and Rodionov 1997).

Similarly to the heat flux, the mass (gas) fluxes vanish at the center.
At the surface, $R$, the gas pressures are those exerted by the coma; in the
lowest approximation they may be assumed to vanish:
$P_{\alpha}(R,t)=0$.
It should be mentioned that the simple (and commonly used) outer boundary conditions
for both energy and gas fluxes have been recently examined in more detail by Davidsson
and Skorov (2002, 2004). Back-scattering of molecules leaving the
nucleus surface and penetration of solar radiation into a thin sub-surface layer of
the nucleus have been shown to affect
the surface and sub-surface
temperatures. These temperatures, however, are equally affected by the other approximations of the model
(such as sphericity or homogeneity of surface structure), not to mention the uncertainty in
the thermal conduction coefficients. Nevertheless, production rates are far {\it less} affected
by these factors. For example, if back scattering reduces the net amount of sublimation at
a given temperature, the surface temperature increases, but with it the sublimation rate
increases as well.
In fact, at low heliocentric distances, the amount of sublimation may be quite accurately
calculated simply by $(1-\Acal)L_{\odot}\cos z/(4\pi d_H^2\Hcal)$, {\it independently}
of the surface temperature.

The numerical code solves the heat diffusion and mass conservation equations
using a fully implicit difference scheme (see Prialnik 1992).
In order to obtain a better resolution near the surface, where the temperature gradient
is steepest, the layers are progressively thinner towards the surface.
With the spin vector perpendicular to the orbital plane, results for 6 hour angles
are recorded: $\varphi=0/360^\circ,\, 60^\circ,\, 120^\circ,\, 180^\circ,\,240^\circ,\, 300^\circ$,
a different one at each time-step.
Separate calculations are carried out for latitudes corresponding to
$\cos\theta=1,\, 0.75,\, 0.5,\, 0.25$,
that is, between the equator ($\theta=0^\circ$) and a near-pole angle ($\theta=75.5^\circ$).
The physical parameter values used in the model calculations are given in Table \ref{tablechar}.

\begin{table}[tbp]
\begin{center}
\caption{Orbital elements and nucleus characteristics}
\smallskip
\begin{tabular}{|l|l|l|l|}
\tableline\tableline
Parameter & Symbol & Value & Units \\
\tableline
 Semi-major axis & $a$ & 3.12 & AU \\
 Eccentricity & $e$ & 0.517491 &  \\
 Effective radius & $R$ & 3.3 & km  \\
 Geometric albedo   & $\Acal$ & 0.04 & \\
 Nucleus spin period & $P_{\rm rot}$ & 41.85 & hr \\
 Dust mass fraction & $X_{\rm d}$ & 0.5 & \\
 Ice mass fraction & $X_{\rm ice}$ & 0.5 & \\
 Porosity & $\Psi$ & 0.5 & \\
 Dust density & $\varrho_{\rm d}$ & 3250 & kg~m$^{-3}$ \\
 H$_2$O ice density & $\varrho_{\rm ice}$ & 917 & kg~m$^{-3}$ \\
 Dust heat capacity & $c_{\rm d}$ & $1.3\times10^3$ & J~kg$^{-1}$~K$^{-1}$ \\
 Ice heat capacity  & $c_{\rm ice}$ & $7.49 T+90$ & J~kg$^{-1}$~K$^{-1}$ \\
 Dust conductivity  & $K_{\rm d}$ & $10$ & J~m$^{-1}$~s$^{-1}$~K$^{-1}$ \\
 C-Ice conductivity   & $K_{\rm c}$ & $5.67\times10^2/T$ & J~m$^{-1}$~s$^{-1}$~K$^{-1}$ \\
 A-Ice diffusivity   & $K_{\rm a}/(c_{\rm ice}\varrho_{\rm ice})$ & $3\times10^{-7}$ & m$^2$~s$^{-1}$ \\
\tableline\tableline
\end{tabular}
\label{tablechar}
\end{center}
\end{table}

\section{Properties of a porous nucleus structure}
\label{newsvr}

\noindent
A cold, icy porous medium allows for sublimation from the pore walls and
condensation onto them, as well as for flow of --- usually dilute --- gas through
the pores. The three fundamental properties that affect these processes are
the porosity $\Psi$, the surface to volume ratio (SVR) $S$ and the permeability
$\phi$. The first is defined as the total volume of pores per given bulk volume,
and the second, as the total interstitial surface area of the pores per given bulk volume.
The permeability is, up to a numerical constant, the proportionality coefficient between
the mass flux and the gradient of $P/\sqrt{T}$ that drives the flow of mass,
where $P$ and $T$ are the gas pressure and temperature, respectively.

In order to determine these three properties, a model of the porous structure is
required.
A commonly used one in modeling cometary interiors is that of a bundle
of unconnected cylindrical tortuous capillary tubes (Mekler et al. 1990), where
the tortuosity $\xi$, defined as the ratio of capillary length to linear distance,
is taken as a free parameter. Another free parameter is the pore (capillary)
radius $r$,
for which some reasonable average pore size is assumed. These assumptions result in
very simple expressions for $S$ and $\phi$ in terms of $\Psi$, $\xi$ and $r$:
\begin{equation}
S={2\Psi\over r}, \qquad \phi={\Psi r\over\xi^2}.
\label{simplesvr}
\end{equation}
These expressions imply, however, that {\it all} capillaries are of the {\it same}
radius. If $r$ represents the mean of a distribution of pore sizes, the simple
formulae above cease to be correct, a fact that has been largely overlooked in
previous calculations. There is, however, a price to pay for a more realistic
approach to pore sizes, and it involves a larger number of parameters for defining
the medium. Nevertheless, we shall show that the deviations from eqs.(\ref{simplesvr})
may be quite considerable and thus worth the price.

Starting from the same model of cylindrical tortuous capillaries, we assume the radii
to vary according to some size distribution, and consider a volume of unit thickness.
Let $N(r)dr$ be the number of capillaries with radii between $r$ and $r+dr$ crossing
a unit area. Keeping in mind that for a capillary tube, $\phi\propto r^3/\xi$ (Gombosi 1994),
the three fundamental properties of a porous medium are, in fact, moments of the
distribution function:
\begin{eqnarray}
S&=&\xi\int 2\pi r N(r)dr \\
\Psi&=&\xi\int \pi r^2 N(r)dr \\
\phi&=&{\pi\over\xi}\int r^3 N(r)dr
\label{moments}
\end{eqnarray}
(cf. Prialnik et al. 2005).
It follows that
\begin{equation}
\phi={\Psi\over\xi^2}{\int r^3 N(r)dr\over\int r^2 N(r)dr}={\Psi\over\xi^2}\bar r
\label{phi}
\end{equation}
while
\begin{equation}
S=2\Psi{\int r N(r)dr\over\int r^2 N(r)dr}= 2\Psi\bar{\left({1\over r}\right)},
\label{svr}
\end{equation}
where $\bar r$ in eq.(\ref{phi}) is the mean pore radius weighted by the volume fraction occupied by
capillaries of radii in the range $(r,r+dr)$. We note that eqs.(\ref{phi}) and (\ref{svr}) are
similar to (\ref{simplesvr}), except that $S$ depends on the harmonic mean of $\left({1\over r}\right)$,
rather than on the inverse of the mean. If we adopt $\bar r$ as the free parameter, then
the expression for $S$ in eq.({\ref{simplesvr}) should be corrected by a factor
\begin{equation}
C=\bar r \bar{\left({1\over r}\right)}={\int r^3 N(r)dr\int r N(r)dr
\over\left(\int r^2 N(r)dr\right)^2}.
\label{correctionfactor}
\end{equation}

A reasonable distribution function for pore size is a power law, since grains ejected from comet nuclei
have a power law size distribution (see below), and comets are believed to be formed by uncompacted
aggregations of grains, in which case the grain and pore size distributions should be
similar. Thus, we assume
\begin{equation}
N(r)\propto r^{-\alpha} \qquad {\rm for} \qquad r_{\rm min}\,<\,r\,<\,r_{\rm max},
\end{equation}
which depends on 4 parameters: the normalization factor, the exponent $\alpha$ and
the range of the distribution defined by $r_{\rm min}$ and $r_{\rm max}$. Instead,
besides $\alpha$, we may choose to assume values for the porosity $\Psi$ (serving as a normalization
factor), the average pore radius $\bar r$ (in order to be able to compare results with the
former, simpler approach), and the ratio
\begin{equation}
X\equiv r_{\rm min}/r_{\rm max}\,<\,1.
\end{equation}
The resulting correction factor for the SVR,
\begin{equation}
C_X(\alpha)={(3-\alpha)^2\over(4-\alpha)(2-\alpha)}{(1-X^{4-\alpha})(1-X^{2-\alpha})
\over(1-X^{3-\alpha})^2}
\label{corfacalpha}
\end{equation}
is plotted in Fig.\ref{correction_factor} for several values of $X$. Following
the distribution of dust grains inferred from observations, typical
parameter values are $X=10^{-3}-10^{-4}$ and $3\le\alpha\le4$ (Fulle et al.
1997, Jockers 1999, Harker et al. 2002).
\begin{figure}[hbtp]
\centering
\scalebox{0.80}{\includegraphics{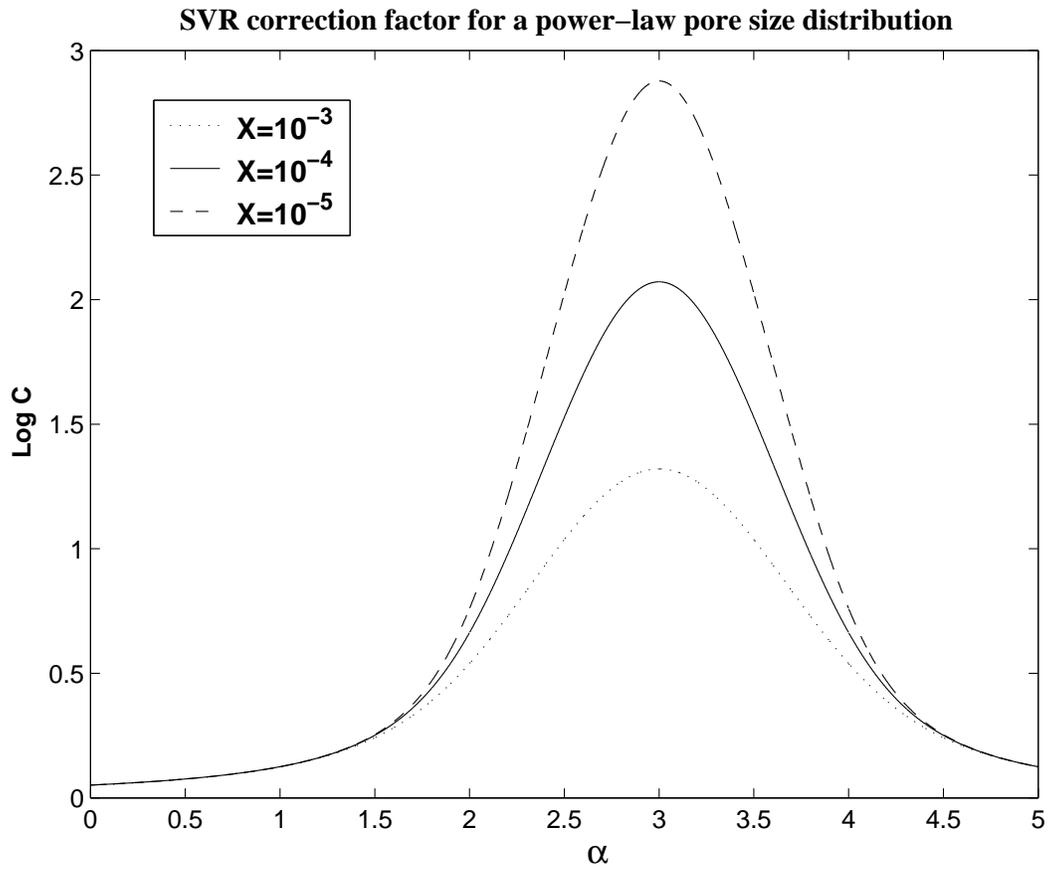}}
\caption{The correction factor of the SVR parameter, as a function of the power law exponent,
for different ratios of minimal to maximal pore size. Note that the interval commonly used
for pore size distribution ($3< \alpha <4$) gives a factor $\ge 10$.}
\label{correction_factor}
\end{figure}

We note, therefore, that for given porosity and
average pore size, taking into account a pore size {\it distribution} can increase
the SVR by up to a factor of 100. Moreover, large corrections correspond precisely
to that narrow range of $\alpha$ values deduced from observations. The effect of a
large SVR should manifest itself both externally,
through the production rates of volatiles, and internally, through composition profiles.
Both are bound to affect, for example, the interpretation of the upcoming {\it Deep Impact}
observations.
A comparison of model results obtained with and without the correction factor for the SVR
will be shown in Section \ref{results}.

\section{New Ground-Based Observations}
\label{observations}

\noindent
In order to prepare for the {\it Deep Impact} encounter, the {\it Deep Impact} team
has undertaken a large observing campaign to characterize the nucleus and
the levels of activity (Meech et al. 2005).  From past apparitions
it was known that this comet typically exhibits a sharp rise in brightness
near 200 days pre-perihelion (Meech et al. 2005).  Optical CCD images have been obtained
regularly since 1997 to monitor the development of activity and the
cessation of activity as the comet moved to aphelion in early 2002.
This comet is well placed for observing every other apparition, and
the early 2000 perihelion passage was only moderately good for dust
dynamical modeling, but was our first chance to measure the dust parameters fot this comet.
Nevertheless, we have a large data set accumulated for this purpose beginning in January
and March 1999 pre-perihelion, and continuing from August 2000 through January 2001.

Finson-Probstein (1968) dust-dynamical models use the observed extent
and morphology of the dust coma to determine the relative velocity
distribution, size distribution and production rates of the dust leaving
the nucleus as a function of heliocentric distance.  These models evaluate the motion
of a suite of particles after leaving the nucleus under the influence of
solar radiation pressure and gravity.  The scattered light from the dust
is added together and fit to the surface brightness of the observed coma.

In order to use dust imaging to constrain the velocity and size distributions
as well as the production rates, the images must be taken at the appropriate
times:

\begin{itemize}

\item Small particles move rapidly away from the nucleus (and out of the
field of view), so in order to constrain the size distribution from models,
many images closely spaced in time need to be obtained.

\item Large grains, which move slowly along the orbit, need observations
equally spaced over time periods of months for proper modeling.

\item Small particles tend to lie in the anti-solar direction, whereas
large particles lag behind in orbit.

\item The predicted dust trajectories (syn-curves, comprising the locus of the synchrones and syndynes)
should be spread out in the plane of the sky so that good constraints can be obtained
for the model parameters.

\item Very circular orbits tend to have widely spaced syn-curves,
but changes with time are harder to see.  Comets on very elliptical
orbits have problems inbound when the large particles fall along the
sun-to-comet radius vector (and overlap the syn-curves for the small
grains).  9P/Tempel 1 is a good comet for modeling in that its orbit
has an intermediate eccentricity.

\end{itemize}

For comet 9P/Tempel 1, the syn-curves were predicted to be moderately
good for the fall of 2000 observations (post-perihelion), and excellent
starting in February 2005 through the encounter.  We will report on the
preliminary modeling for the observations obtained during fall 2000.

Kron-Cousins R-band images were obtained on 2000 Aug 21 and 2000 Sep 30,
using the University of Hawaii 2.2m telescope on Mauna Kea and the the
Tektronics 2048$\times$2048 CCD
(read noise = 7e$^{-}$, gain = 1.4e$^{-}$/ADU, plate scale =
0.219$\farcs$~pixel$^{-1}$). Flat fields were obtained on
the twilight sky, and standard reduction procedures were able to flatten
the data to with 0.4\% across the CCD.  The nights were photometric,
and standard stars from Landolt (1992) were observed on each night.

Composite
images were made by using measurements of the centroids of a large
number of field stars  and the centroids were used
to compute the offsets between the individual images (from telescope
guiding errors).  After the offsets were applied, the ephemeris rates for
9P/Tempel 1 were used, in combination with the image plate scale, to
calculate shifts to simulate guiding at non-sidereal rates, and then
the images were added. Selected field stars, close to the comet, were fit and
removed from composite image.
The resultant images are shown in Figure xx (attached to the paper).

\noindent
Fig.xx.-- Composite images of comet 9P/Tempel 1 (left)
from 2000 Aug. 21 ($r$=2.54 AU), composed of 26 $\times$ 300 sec R
images and (right) from 2000 Sep. 30 ($r$=2.77 AU) composed of 42
$\times$ 300 sec R images. Images are 180$\times$180$''$, with N at
the top and E to the left.


Using a code developed by Farnham (1996), Finson-Probstein
dust dynamical models were fit to the images, and preliminary results
are shown in Table~\ref{tab.FPmodel}. Contour fits to the images are shown
in Figure~\ref{karen2}.

\begin{table}[h]
\begin{center}
\caption{\label{tab.FPmodel} Finson-Probstein Model Fit}
\smallskip
\begin{tabular}{ll}
\tableline\tableline
{\bf Parameter} & {\bf Value} \\
\tableline
Smallest particles            & 3 $\mu$m                                  \\
Largest particles             & 3 mm                                      \\
Emission start                & $perihelion$ -- 100 days                           \\
Maximum dust output           & $perihelion$ -- 60 days                            \\
Emission decline              & $perihelion$ + 260 days                            \\
Velocity $(v(\beta))^{[1]}$   & $v(r_{\rm min})/v(r_{\rm max})\approx40$  \\
$\alpha$$^{[2]}$              & 3.1                                       \\
\tableline
\end{tabular}
\end{center}
Notes: $^{[1]}\beta=F_{rad}/F_{grav} = 5.74\times10^{-4}Q_{rp}/\varrho_{d}r$,
is a proxy for grain size, where $Q$ is the radiation pressure scattering efficiency.
$^{[2]}$ Fit value for the slope of the
particle size distribution.
\end{table}

\begin{figure}[hbtp]
\centering
\scalebox{0.70}{\includegraphics{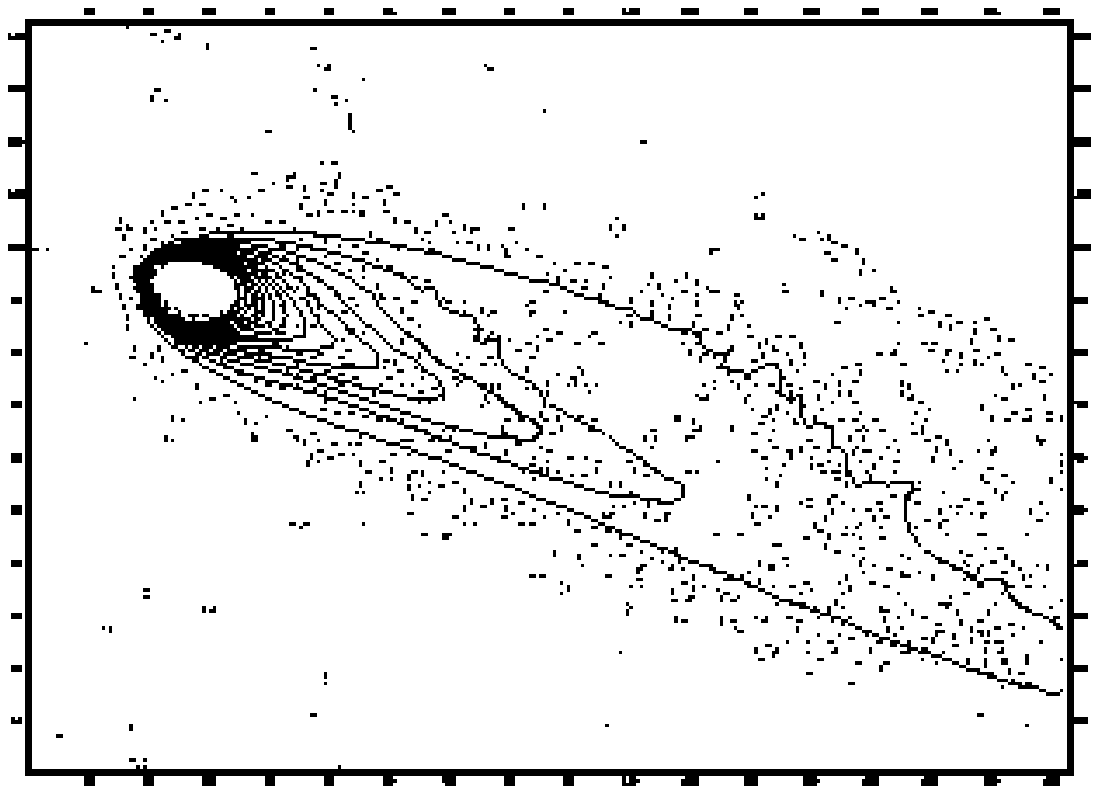}\quad\includegraphics{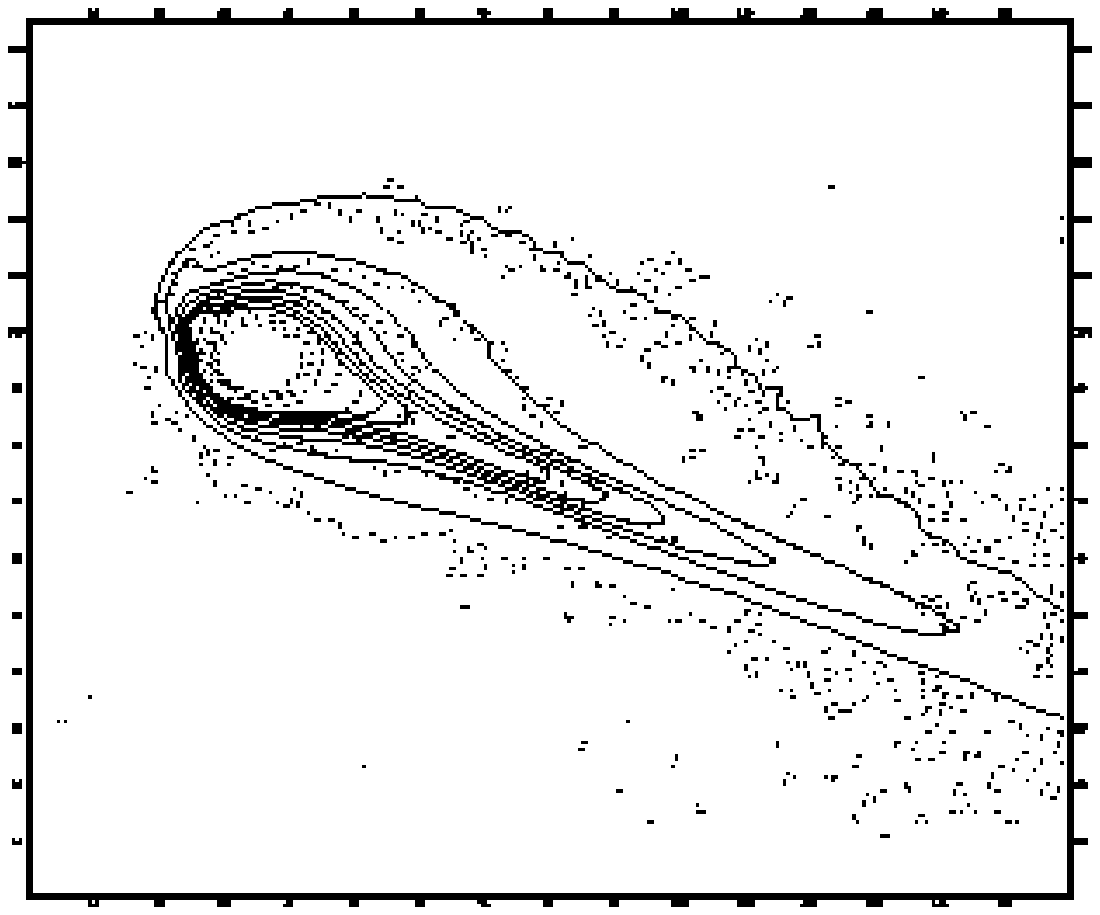}}
\caption{Finson-Probstein dust dynamical model fits to the data on
Aug. 21 (left) and Sep. 30 (right), 2000 plotted on contours from the composite images.}
\label{karen2}
\end{figure}

\section{Results of evolutionary calculations}
\label{results}

\noindent
Since little is known about any comet's structure and composition,
our first task was to choose the structural and compositional
parameters. The composition was chosen by running several models of various
generic composition, but identical structure --- as listed in Table \ref{tableinit} --- and
comparing the resulting production rates to available observations
(Cochran et al. 1992, Osip et al. 1992).
This is shown in Fig.\ref{3specie_comp}. The model for which
agreement was obtained for all observed species available
was then chosen as the {\it working model}.
The dust production rate of this model, shown in Fig.\ref{dustprod}
agrees very well with the results
derived from ground-based observations described in Section~\ref{observations}.
We note, in particular, the times of rise, maximum and decline in the rate
of dust emission, compared to those of Table~\ref{tab.FPmodel}.
The magnitude corresponding to the dust production rate at maximum is
about 9; varying the average grain size and/or dust grain density may
result in a change of $\pm0.5$~mag (for a correlation between dust production
and magnitude, and perihelion magnitudes, see Meech et al. 2005, 2002).
\begin{table}[tbp]
\begin{center}
\caption{Initial composition of 9P/Tempel1 models}
\smallskip
\begin{tabular}{|l|l|l|l|l|}
\tableline\tableline
Constituent & Working model: & Trapped gas & Ice mixture & Ice mixture \\
mass & Trapped gas & Dust mantle & No mantle & Dust mantle \\
fraction & No mantle & &  & \\
\tableline
$X_{a}$   & 0.5 & 0.5 & 0 & 0 \\
$X_{c}$   & 0   & 0 & 0.45 & 0.45 \\
$X_{d}$ & 0.5 & 0.5 & 0.5 & 0.5 \\
$X_{CO}$   & 0.05 & 0.05 & 0.025 & 0.025 \\
$X_{CO_2}$ & 0.0125 & 0.0125 & 0.00625 & 0.00625 \\
$X_{HCN}$  & 0.0035 & 0.0125 & 0.00625 & 0.00625 \\
$X_{NH_3}$ & 0.0125 & 0.0125 & 0.00625 & 0.00625 \\
$X_{C_2H_2}$ & 0.0012 & 0.0125 & 0.00625 & 0.00625 \\
\tableline\tableline
\end{tabular}
\label{tableinit}
\end{center}
\end{table}

\begin{figure}[hbtp]
\centering \scalebox{0.80}{\includegraphics{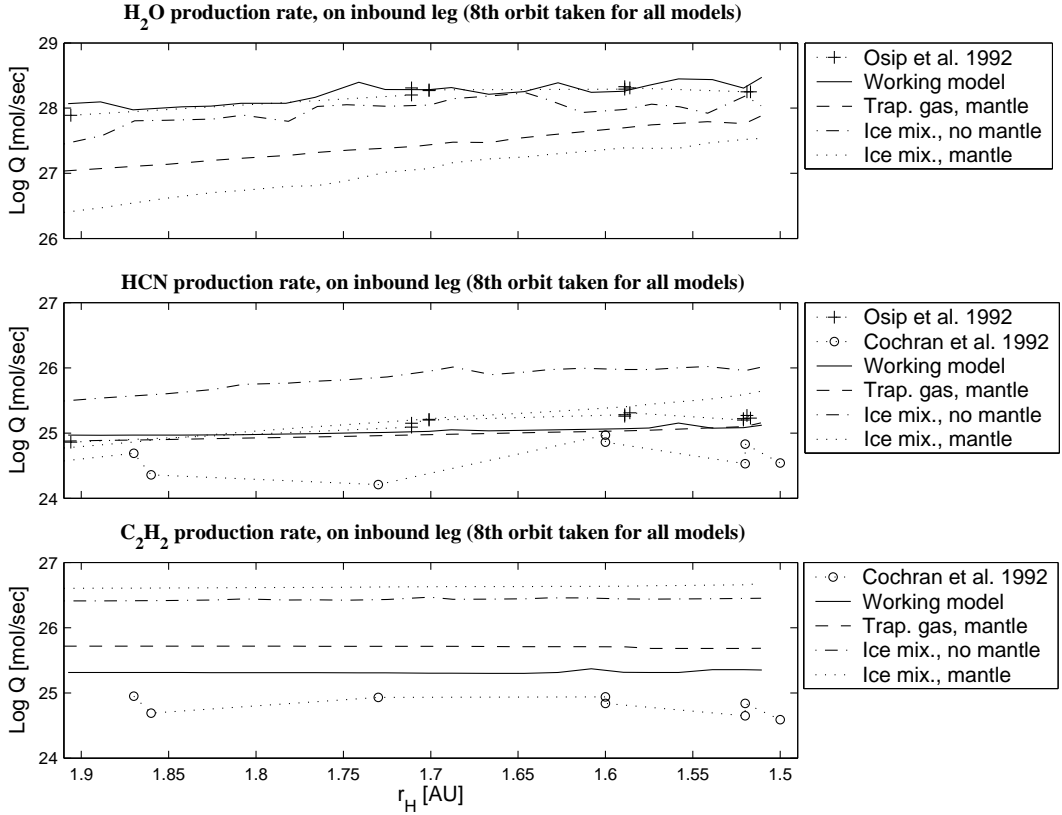}}
\caption{Comparison of observation and model run results for 4
generic models, as described in the paper. Observed production rate
of OH, as the product of H$_2$O, CN, as the product of HCN, and
C$_2$, as the product of C$_2$H$_2$, were taken from the
references.}
\label{3specie_comp}
\end{figure}

\begin{figure}[hbtp]
\centering \scalebox{0.60}{\includegraphics{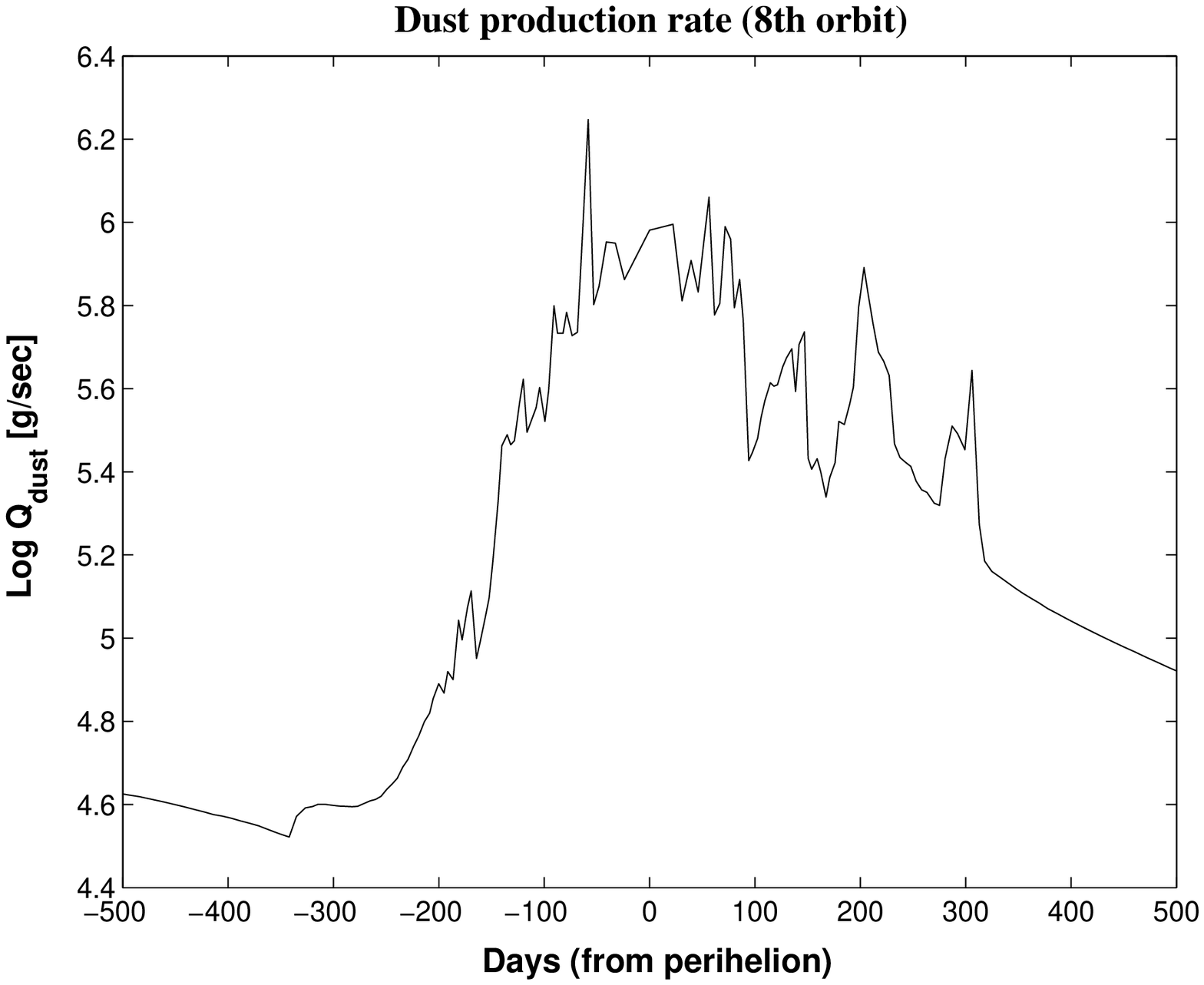}}
\caption{Dust production rate pre- and post-perihelion for the baseline
model.}
\label{dustprod}
\end{figure}

Guided by the dust grain size distribution derived from observations, we adopted for the pore
size distribution inside the nucleus a slightly steeper power law, with $\alpha=3.5$,
an average pore size of 100~$\mu$m, and a ratio $X=10^{-4}$. According to the discussion in
Section \ref{newsvr}, $\bar r\approx\sqrt{r_{\rm min}r_{\rm max}}$, which results in a size
range between $r_{\rm min}=$1~$\mu$m  and $r_{\rm max}=$1~cm. The range was chosen to be wider
than the dust grain size range (3~$\mu$m--3~mm, see Table \ref{tab.FPmodel}) in order to account
for micro-pores at the low-end, and to enable the flow of a few mm-size grains requiring somewhat larger
pores at the high-end.
A steeper power law was chosen for pore sizes than indicated by grain observations, because dust
grains are also lifted off the surface, where pore size does not constitute an impediment, and hence
the relative proportion of larger grains coming from the surface should be larger than for those
originating from the interior.

Assuming the spin axis to be perpendicular to the orbital plane and
starting with a uniform low temperature, we followed the evolution
of the working model for several orbits. The thermal and compositional
evolution is illustrated in Fig.\ref{T_Aice_prof}, for the subsolar
point on the equator and for a point near the pole. These represent extreme cases --- maximum
and minimum --- with respect to absorption of solar radiation, and hence activity, of a
spinning nucleus in the orbit of 9P/Tempel~1, among {\it all} possible inclinations of the spin axis.
We recall that the model nucleus is spherical, whereas in reality the nucleus is found to be
highly elongated, with semi-axes $a\approx7.2$~km, $b\approx c\approx 2.2$~km (Belton et al. 2005).
Thus the "collecting area" of the nucleus may vary, according to inclination, between
$\sim15\,-\,50$~km$^2$, compared to 34~km$^2$ of the model. This would place an error bar of up to
a factor of 2 on the results regarding production rates, since most of the solar heat is spent in
sublimation of volatiles.
The shape itself should be of lesser importance, since the affected
layer, as we shall see, is much smaller than the radius.

The outstanding feature emerging from this calculation is the complicated
stratification pattern as a function of depth, where layers enriched in various volatiles
alternate. Moreover, several layers enriched in the same volatile may appear at different
depths.
The effect is illustrated by the mass fraction
of amorphous ice shown in the right panels of Fig.\ref{T_Aice_prof}.
We recall that the model's composition includes 5 volatile species trapped in
amorphous water ice. These volatiles cover a wide range of sublimation temperatures
(see Prialnik et al. 2005). As the surface of the nucleus is heated and the heat wave
propagates inward, the amorphous ice crystallizes and the trapped gas is released.
The gas pressure in the pores peaks at the crystallization front. As a result,
gas flows in part outward and escapes, and in part, inward into colder regions.
Eventually, each species reaches a sufficiently cold region for it to recondense.
Since recondensation releases heat, it affects the composition of its surroundings
and thus a complicated pattern results, of alternating ices mixed with the
amorphous water ice. When another heat wave reaches these regions, on a subsequent
perihelion passage, the heat is absorbed in sublimation of the recondensed volatiles
rather than in crystallization of amorphous ice.
This is how alternating layers of
crystalline and amorphous ice arise, rather than a single boundary between a
crystalline exterior and an amorphous interior.
Finally, the {\it steps} that appear in the outer crystalline/amorphous ice boundary
are due to erosion of the
nucleus, which brings this boundary closer to the surface.

The stratified layer extends from a depth of about 10~m below the suface and down to a few
hundred meters. Since 10~m is roughly the orbital skin depth for 9P/Tempel~1, this structure
should cause activity variations on the orbital time scale. This means that activity may
differ from orbit to
orbit and occasional spurious outbursts may arise, when the heat wave propagating down from the surface
reaches a region enriched in ice of some volatile species. Such outbursts of gas should be
accompanied by ejection of dust. Indeed, this variable behavior is exhibited in Fig.\ref{SVRcomp}, where we
note an outburst of gas (CO and CO$_2$) and dust following several "quiet" regular orbits. By contrast,
water production, whose main source is sublimation from the surface, follows a much more regular pattern.

The evolution of local noon temperature profiles is shown in the left panels of Fig.\ref{T_Aice_prof}.
It indicates that a steady state is reached in each case after a few revolutions
at about the same depth, both near the pole and at the equator. This depth corresponds to the orbital
skin depth $[2Ka^{3/2}/(\sqrt{GM_\odot}\rho c)]^{1/2}\approx10$~m. We note, however,
that the maximum temperatures at the two
locations differ by about 40~K; accordingly, the skin depth, which is roughly inversely
proportional to temperature, is slightly larger near the pole. Without the correction factor
to the SVR, internal temperatures are somewhat higher in the outer layers since less heat is
absorbed in ice sublimation.

\begin{figure}[hbtp]
\centering
\scalebox{0.35}{\includegraphics{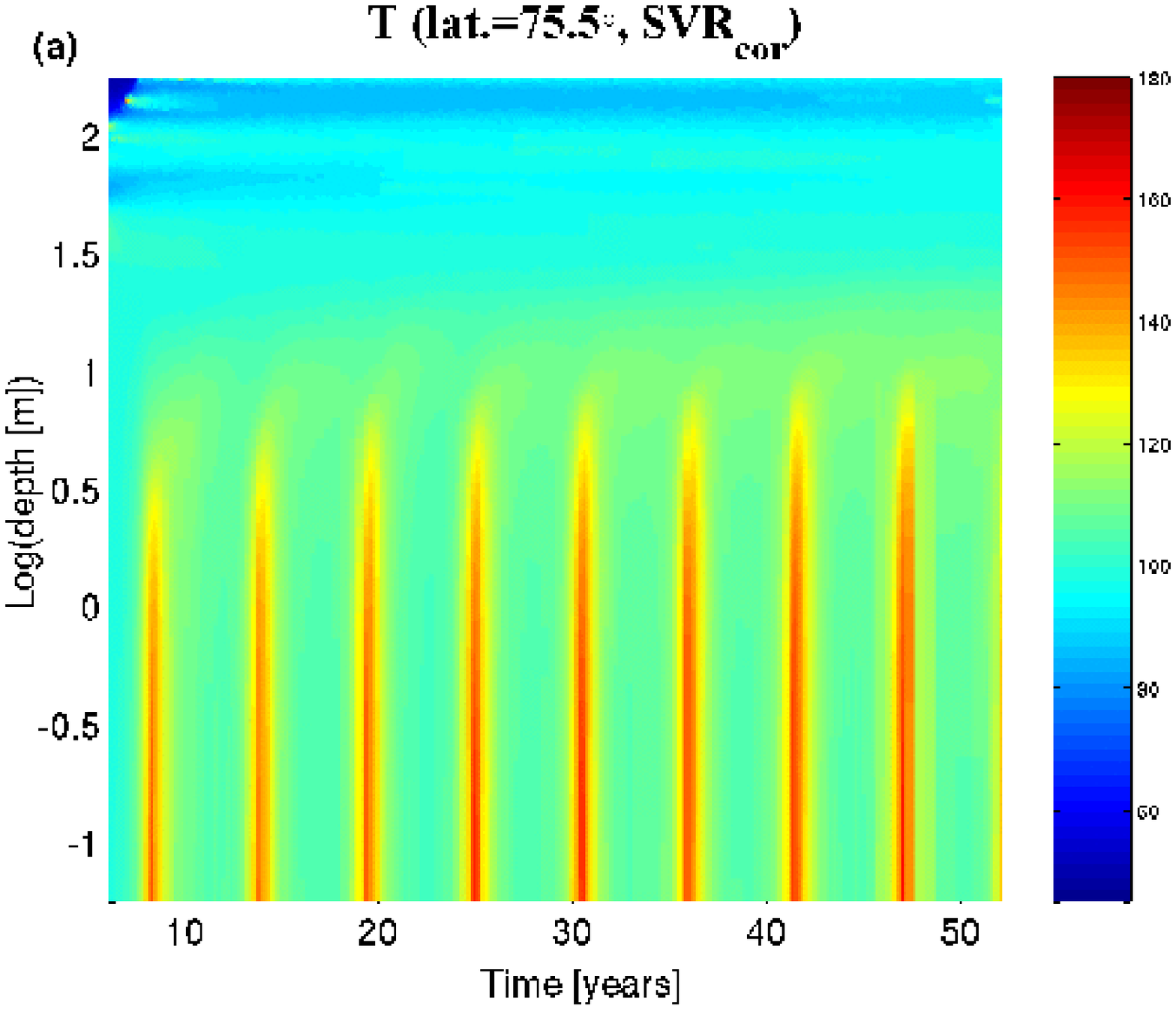} \includegraphics{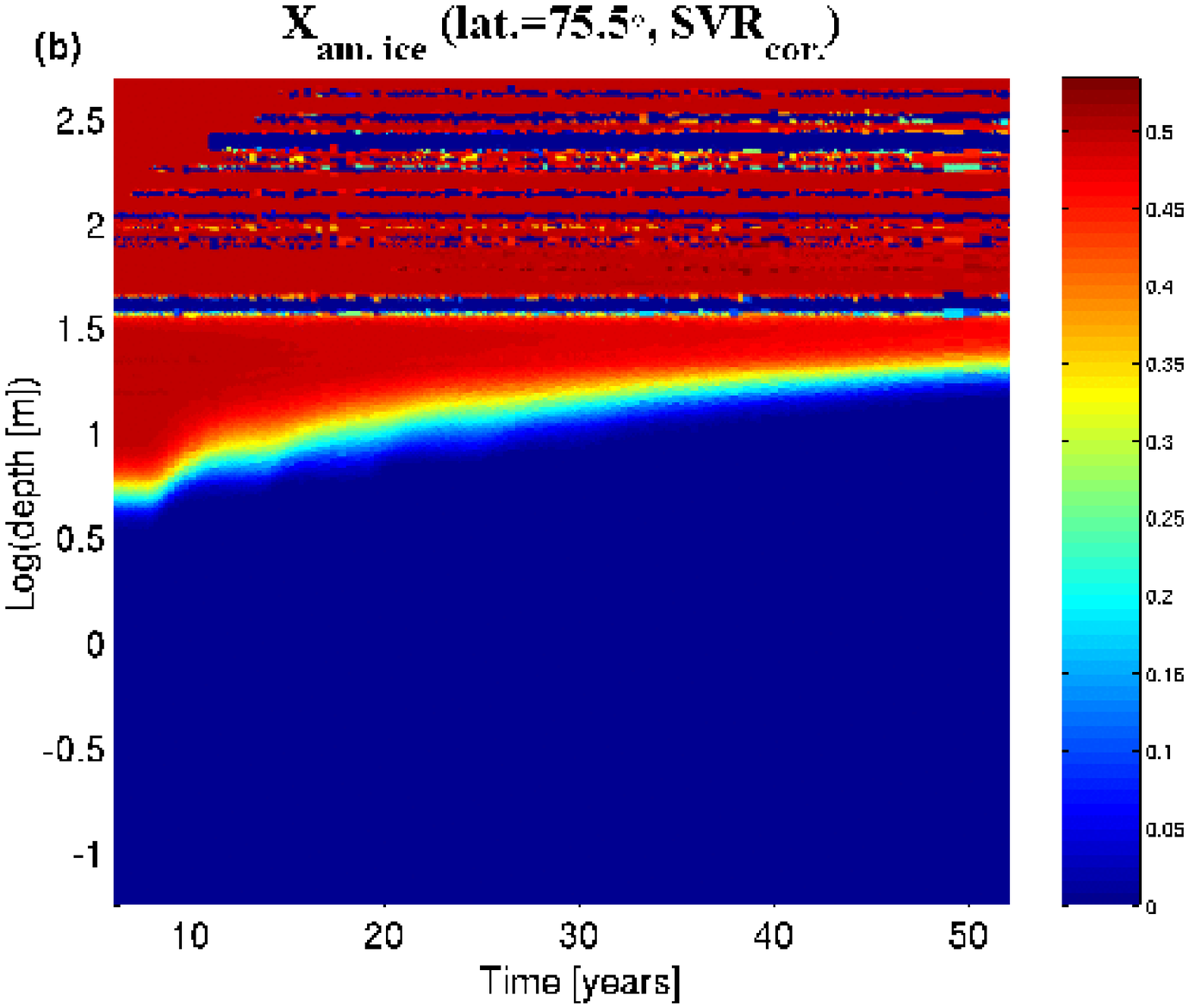}}
\scalebox{0.35}{\includegraphics{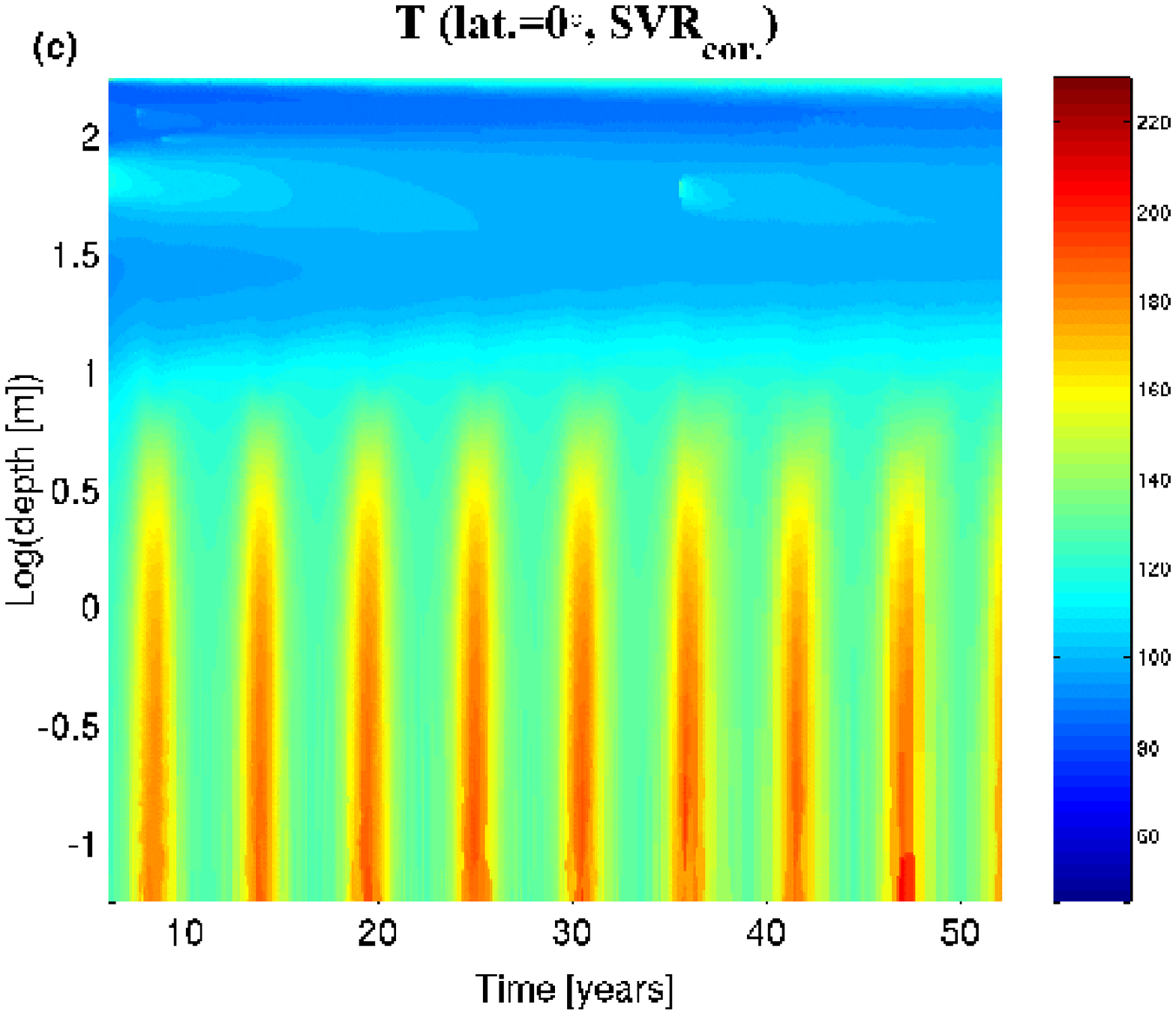} \includegraphics{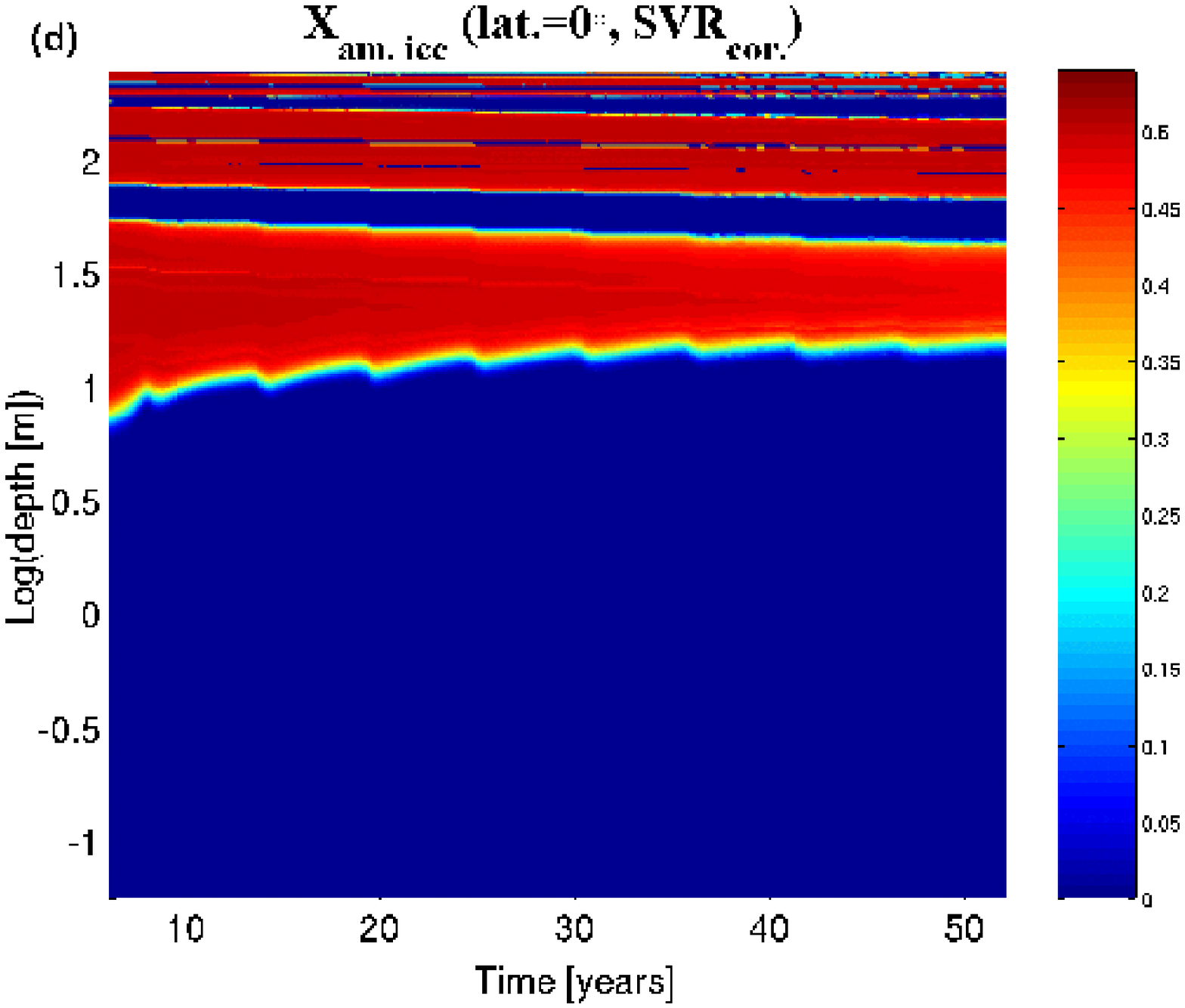}}
\scalebox{0.35}{\includegraphics{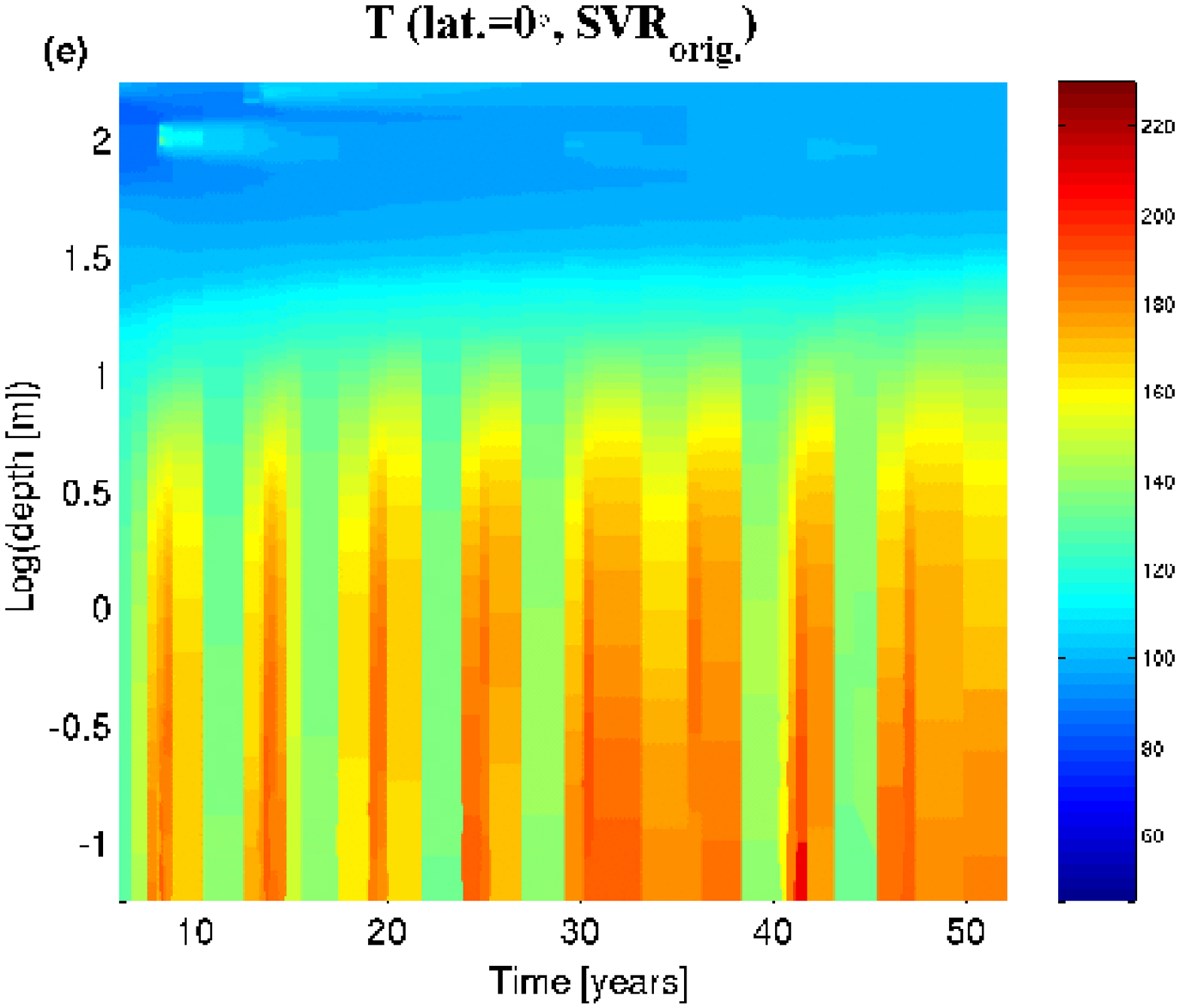} \includegraphics{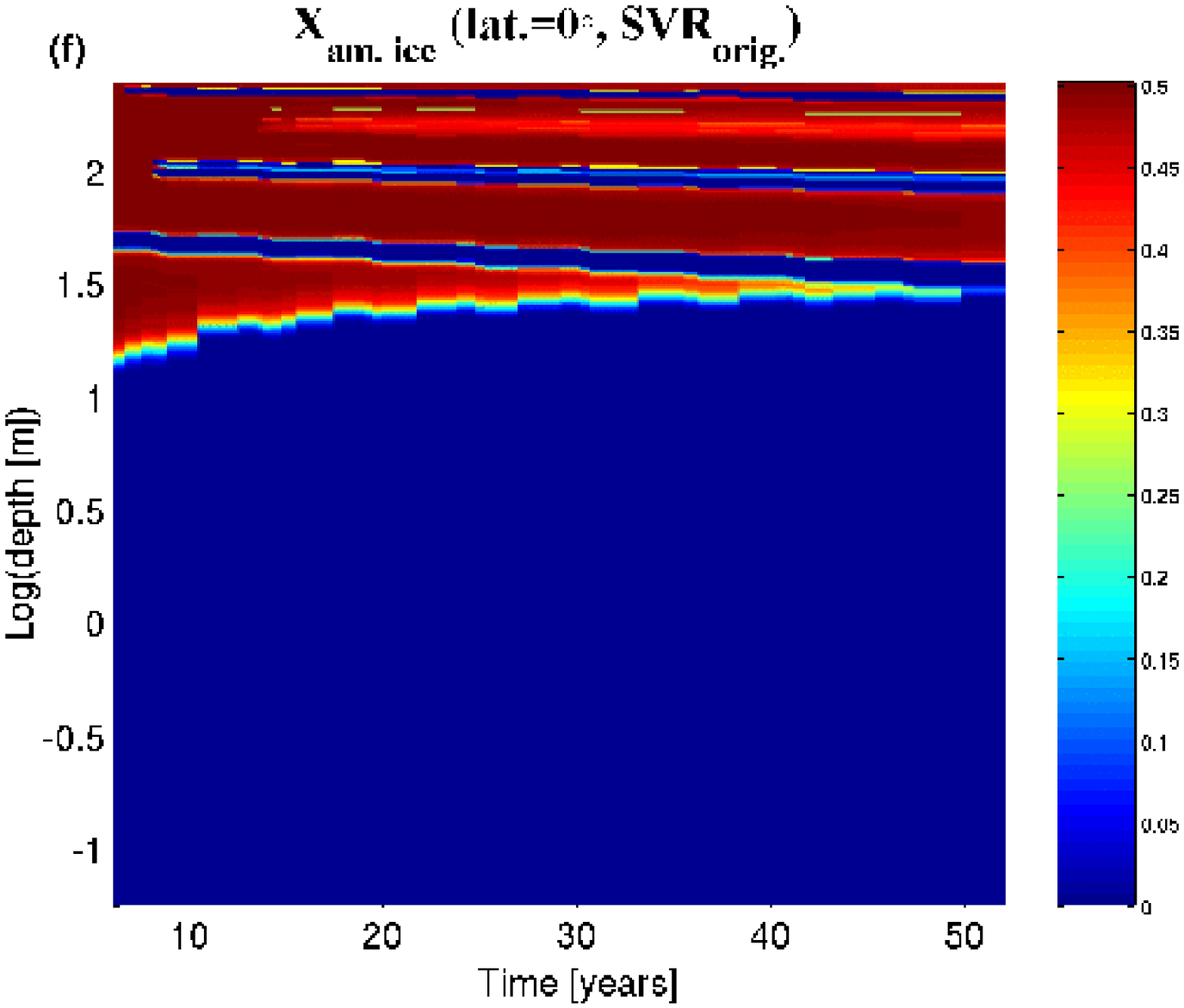}}
\caption{Evolution of the comet nucleus working model through repeated revolutions. The
temperature and amorphous ice abundance are shown as a function of time and depth:
(a) and (b) --- near-pole latitude, with the corrected SVR term;
(c) and (d) --- equator, with the corrected SVR term;
(e) and (f) --- equator, with the original SVR term.
The depth scale is logarithmic, from the surface to the interior.}
\label{T_Aice_prof}
\end{figure}

We now wish to draw attention to the effect of the corrected SVR on the results.
Temperature profiles at the equator are compared in Fig.\ref{T_Aice_prof} in the two
lower left panels, and the evolution of crystallization in the two lower right panels.
The correction factor increases the internal pore surface and enhances
sublimation from the pore walls. This hinders the heat wave from penetrating
deeper and internal temperatures are thus lower. As a result, the depth of
amorphous ice is shallower, as illustrated in Fig.\ref{T_Aice_prof}.
At higher latitudes, due to the diminished insolation, it is still shallower.

Production rates obtained for long-term evolution with and without the SVR correction
are compared in Fig.\ref{SVRcomp}. As expected, surface properties, such as temperature and
H$_2$O production rate, are not affected. The production rates of gases released in the interior,
either escape from crystallizing ice or by sublimation of recondensed ice, are affected to a
larger extent, but preserve the same pattern of behavior. This is not surprising,
since the driving energy source is the same. We note that the complex stratified structure of
the outer layer of the nucleus gives rise to occasional outbursts. Clearly, the orbital
evolution of production rates differs considerably among different volatiles. This means that
observed volatile abundances do not reflect nucleus abundances (cf. Huebner and Benkhoff 1999,
Prialnik et al. 2005).

\begin{figure}[hbtp]
\centering
\scalebox{0.29}{\includegraphics{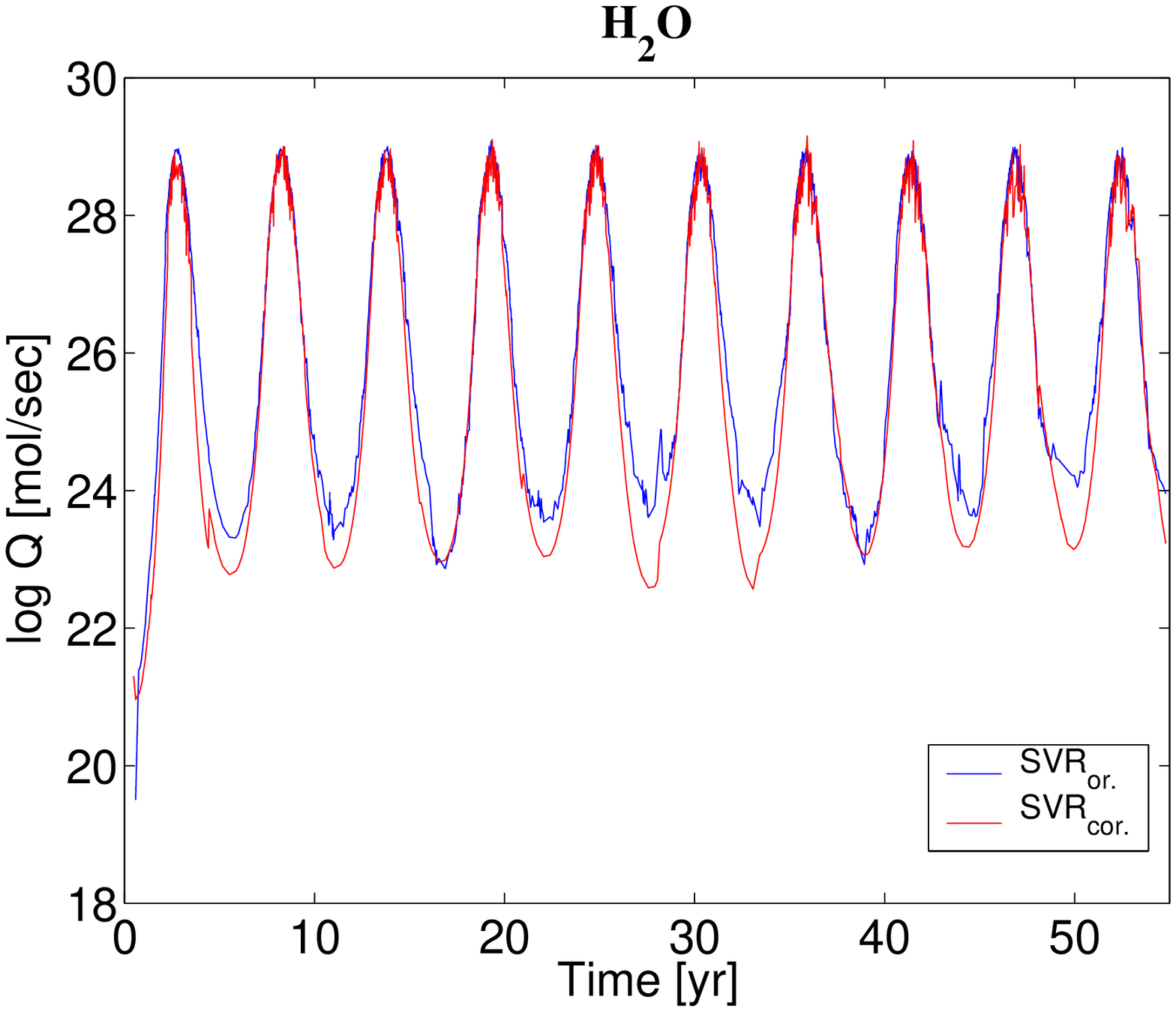} \qquad \includegraphics{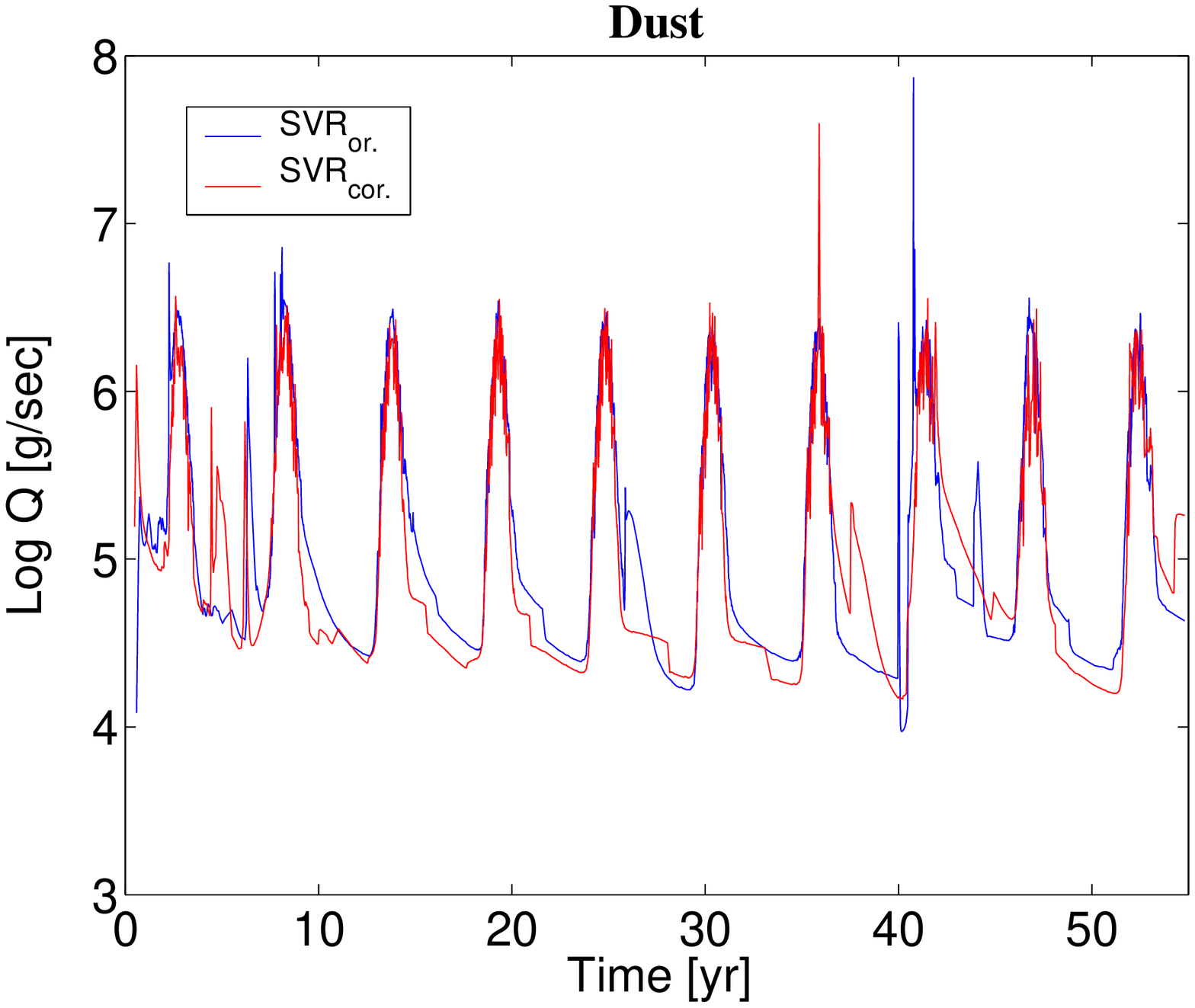}}
\scalebox{0.29}{\includegraphics{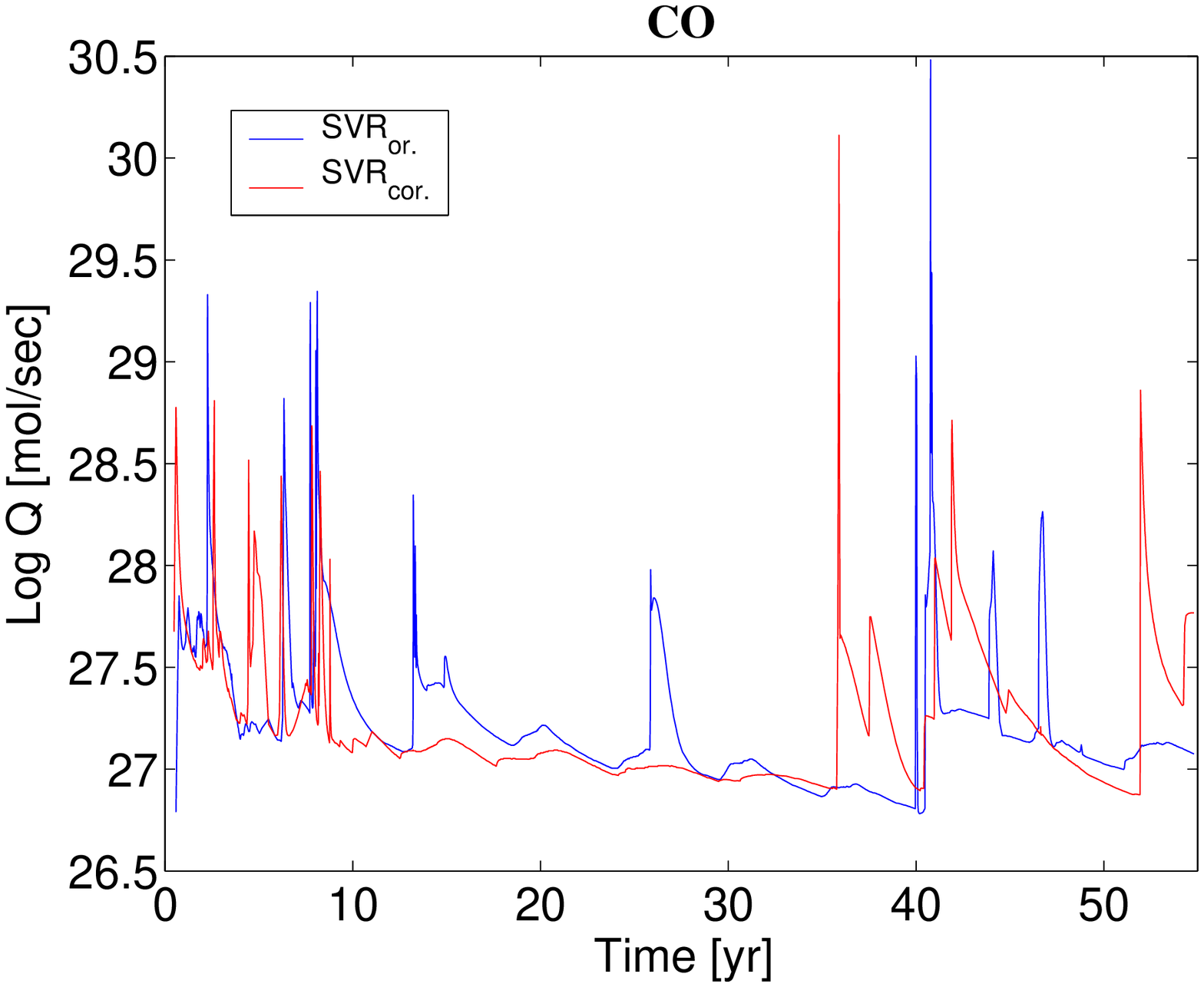} \qquad \includegraphics{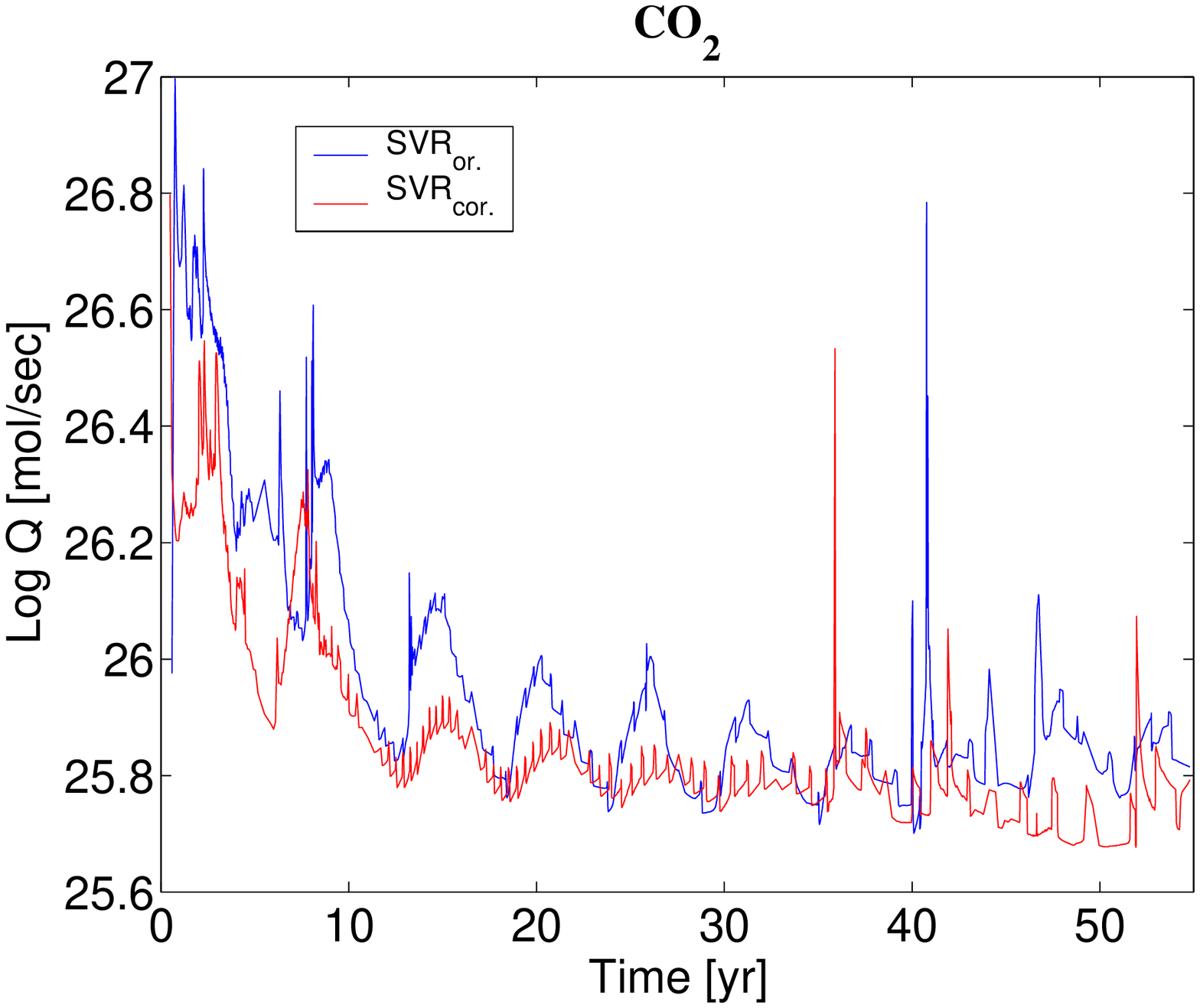}}
\scalebox{0.29}{\includegraphics{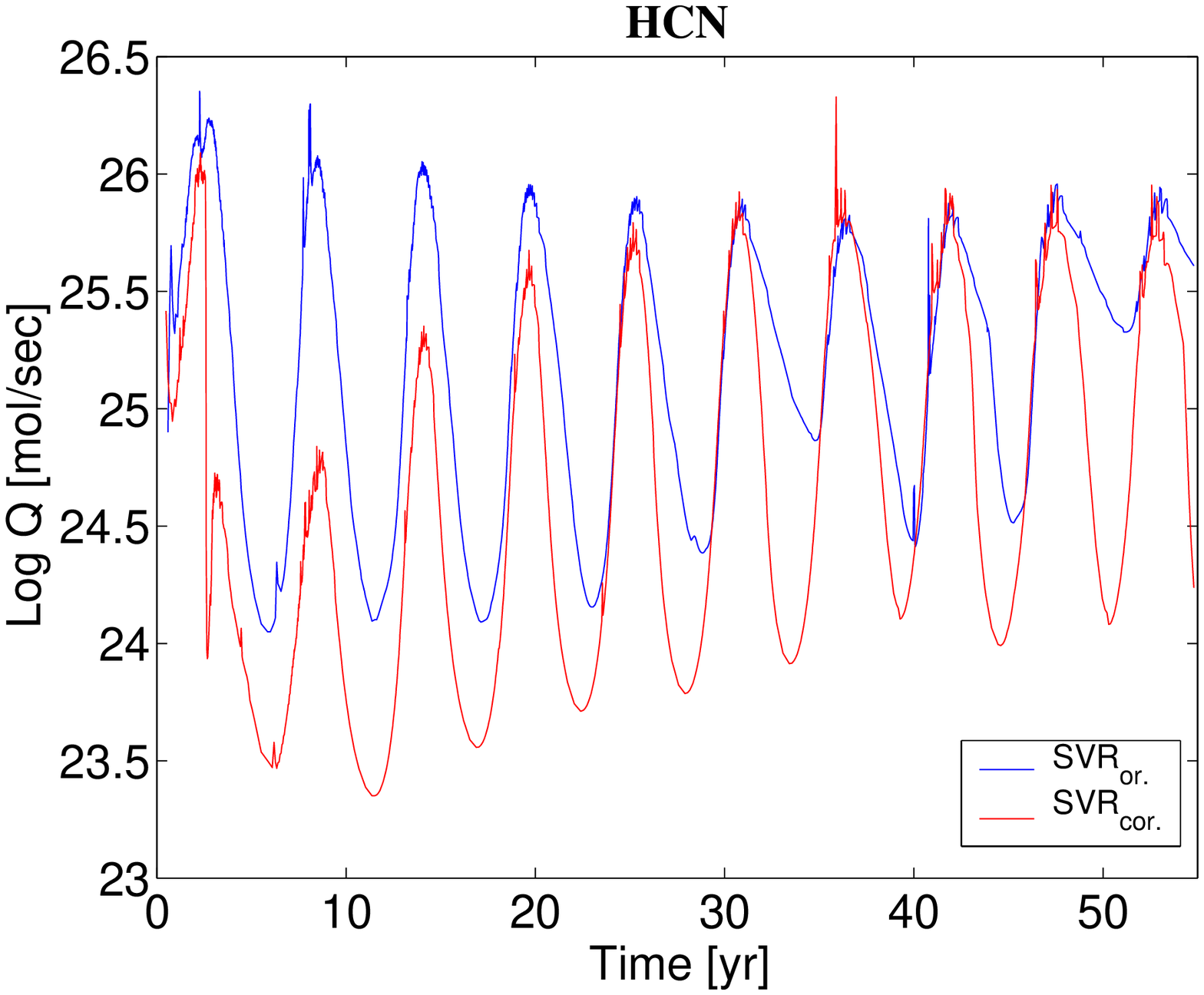} \qquad \includegraphics{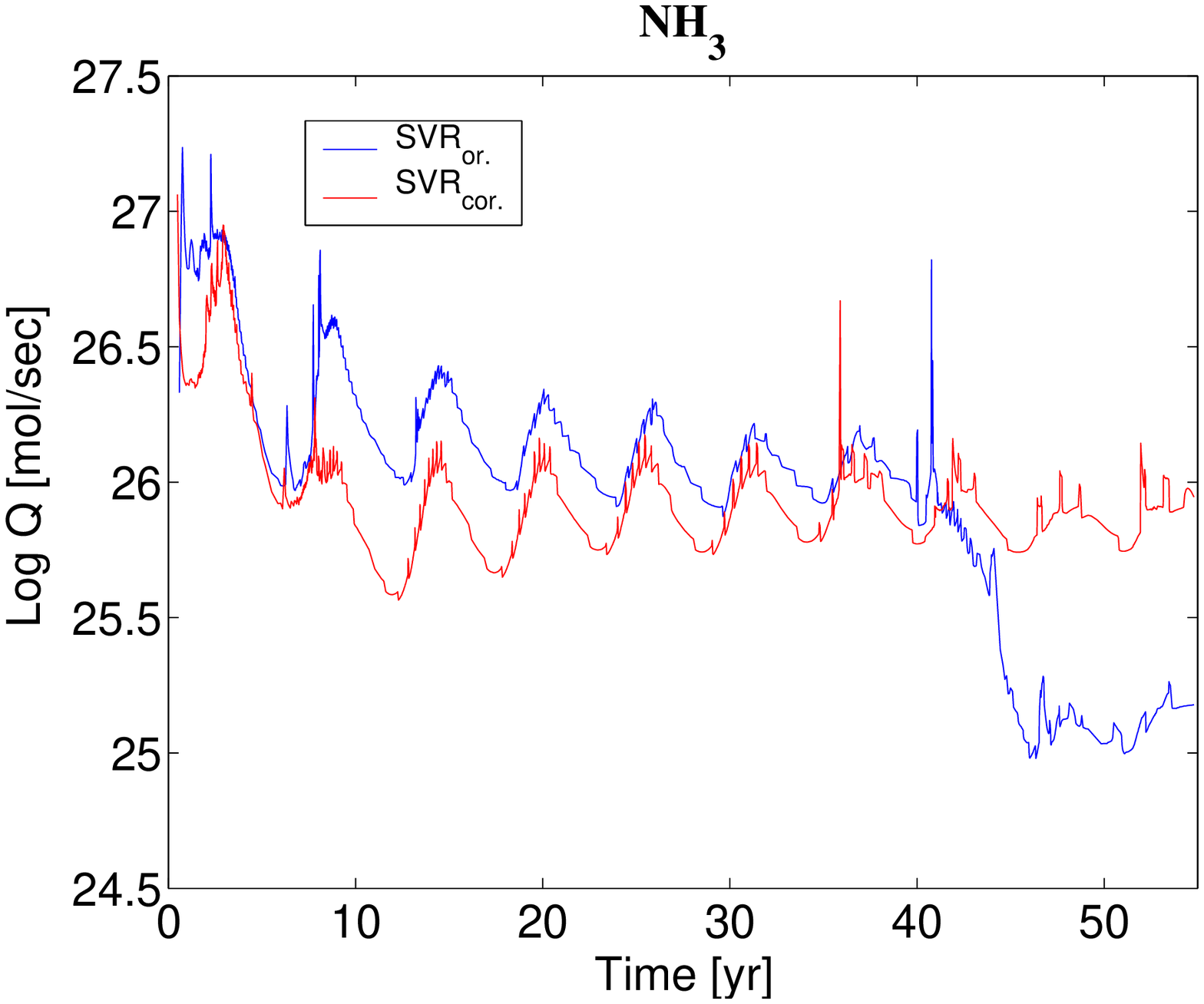}}
\scalebox{0.29}{\includegraphics{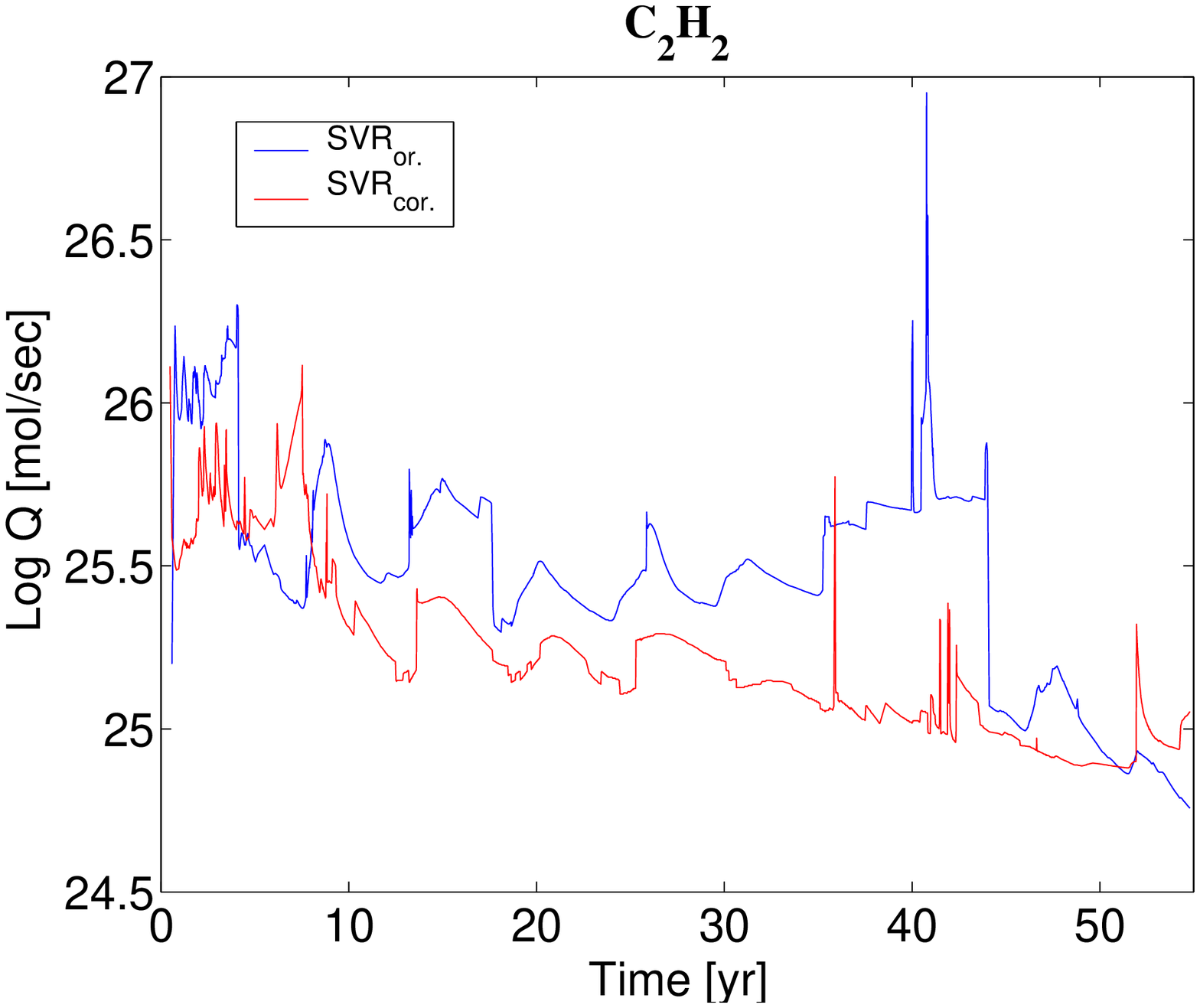} \qquad \includegraphics{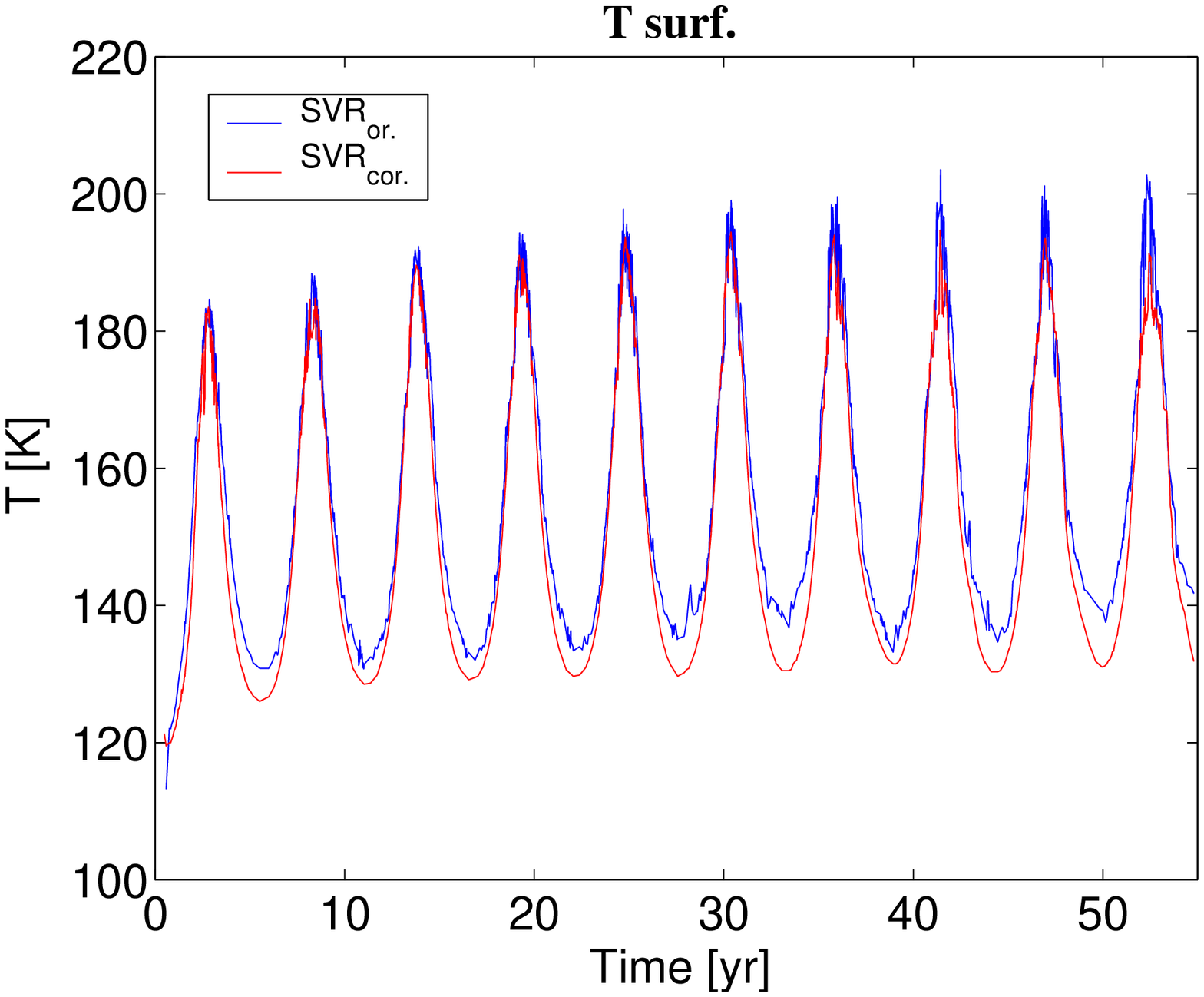}}
\caption{Comparison of original SVR (blue) and corrected SVR (red), for 10 orbits of the working model:
results are shown for the surface temperature at the subsolar point as well as
production rates for all volatiles and dust.}
\label{SVRcomp}
\end{figure}

\section{Simulation of a deep impact}
\label{simpact}

\noindent
The aim of the {\it Deep Impact} mission is the collision of the
impactor spacecraft with the nucleus, in order to create a crater.
This will enable the flyby spacecraft to perform observations and
measurements of the impact itself, the ejected material and the
interior composition, as revealed by the exposed crater.

The thermal effect of the impact is simulated by assuming an
additional energy influx for an infinitesimal period of time. The
known input parameters for the impact simulation are the total
(kinetic) energy of the projectile $E_{\rm tot}$, the estimated area
$A$, that will be affected by the impact --- which is needed in
order to obtain an energy flux --- and the heliocentric distance
(preperihelion) where the impact is expected to take place, $d_0$.
The total energy absorbed per unit area is $E_A=E_{\rm tot}/A$.
We note that the relevant parameter is $E_A$, hence the calculation
is not sensitive to the total impact energy or the area, separately.

\begin{table}[htbp]
\begin{center}
\caption{Impact simulation properties}
\smallskip
\begin{tabular}{|l|l|l|l|}
\tableline\tableline
Parameter & Symbol & Value & Units \\
\tableline
 Total energy & $E_{\rm tot}$ & $1.9\times10^{10}$ & J \\
 Crater area  & $A$ & 1000 & m$^2$ \\
 Crater depth & $\Delta L$ & 30 & m \\
 Heliocentric distance & $d_0$ & 2 & AU \\
 Revolutions before impact & $n$ & 7 & \\
 Timescale & $\tau$ & 180 & s \\
 Peak energy flux & $F_0$ &$10^5$& J~m$^{-2}$~s$^{-1}$ \\
\tableline\tableline
\end{tabular}
\label{tabledi}
\end{center}
\end{table}

Since numerical calculations do not allow singularities, we assume
the energy deposition rate to be in the form of a narrow
time-dependent Gaussian (rather than a delta function), {\it centered} at
the time of the impact, with a width corresponding to the timescale
for the formation of the crater,
\begin{equation}
F(t)=F_0{\rm e}^{-\pi[(t-t_0)/\tau]^2}.
\end{equation}
Here $\tau$ is a free numerical parameter that may be interpreted as the
impact timescale. Normalizing the deposition rate,
\begin{equation}
\int_{-\infty}^{\infty}F(t)dt\,=\,E_A
\end{equation}
we obtain $F_0=E_A/\tau$. Thus, the amplitude of the energy
deposition rate is related to the physical parameters of the impact.
The parameter values used are listed in Table~\ref{tabledi}.
The resulting peak energy flux is about 300 times higher than the solar energy
flux at that distance at the subsolar point. At a distance of 1.5~AU, the current
impact distance, the contrast would be reduced by a factor of $\sim0.6$, but this
change should not be significant.
The time of impact $t_0$ is calculated by
\begin{eqnarray}
t_0&=&\sqrt{a^3/GM_\odot}(2\pi n-\pi-\vartheta_0+e\sin \vartheta_o) \\
\vartheta_0&=&\cos^{-1}[(1-d_0/a)/e],
\end{eqnarray}
where $n$ is the number of orbital revolutions calculated prior to the impact.

At the onset of the impact --- defined to occur at a time $t_0-\nu\tau$, where $\nu>1$, so
that $F(t\pm \nu\tau)\ll F_0$ --- we remove from the model nucleus a
layer of thickness $\Delta L\approx 30$~m, to simulate the crater
that is expected to form. Thus the thermal energy is deposited below
the original nucleus surface, where the composition had been either
preserved, or altered by earlier evolution to a different extent. We
should bear in mind that the thermal effect of the impact is
maximized, since it is assumed that the entire kinetic energy is
turned into heat and the heat is all deposited at a depth
corresponding to the bottom of the impact crater.

Two impact calculations were carried out for two different locations on the
nucleus surface: one at the equator, and another near the pole. The purpose was to find
out to what extent is the point of impact expected to affect the outcome.
The comparison of the two locations is presented in Fig.\ref{eqpolcomp}, and shows that,
to within a fraction of a magnitude, the production rates of the major constituents
(H$_2$O, dust, CO and CO$_2$) are indistinguishable at the moment of impact. However,
since the flyby spacecraft is supposed to conduct the observation for a period of $\sim15$
minutes after the impact, the emission of dust, CO and CO$_2$ may vary by $\sim2$
orders of magnitude during this time.

The low production rates before impact are an artifact of our simulation: when we remove
an outer layer of several tens of meters, we expose cold material. Since the energy is
supplied slowly at first, production rates drop. The total amount of dust ejected upon
impact amounts to $\sim1.7\times10^5$~tons at the near-pole location, and $\sim2.2\times10^5$~tons
near the equator. Compared with the dust production rate over the entire comet at
perihelion, these amounts are equivalent to the total dust output of 2~days and 2.6~days,
respectively, at perihelion.

We note that the CO and CO$_2$ ejection rates are not smooth: they show two spikes and
a somewhat later peak (see Fig.\ref{eqpolcomp}). This is due to the layered structure of
the nucleus described in Section~\ref{results}. The fact that the behavior pattern of these volatiles
is reflected in the rate of dust ejection indicates that a significant fraction of the dust
originates in layers beneath the surface. Small dust grains are thus dragged by volatiles
through the pores. This effect is less marked near the pole, where the subsurface layers
have been processed to a far lesser extent during evolution prior to the impact.

An interesting result of the long-term evolution, is that even though the thermal
energy influx due to the impact is confined in both time (by the Gaussian function)
and space (by the depth of penetration), the effect lingers and shows deviations in
production rates  - as compared with the undisturbed model - long after the event.
However, this effect cannot be used to characterize a comet that has undergone an "impact",
because the evolution pattern is similar to that of a comet model that was not disturbed.
The difference is in the detailed variation of production rates with heliocentric distance,
but these details depend to a similar extent on model parameters and assumptions.

\begin{figure}[hbtp]
\centering \scalebox{0.70}{\includegraphics{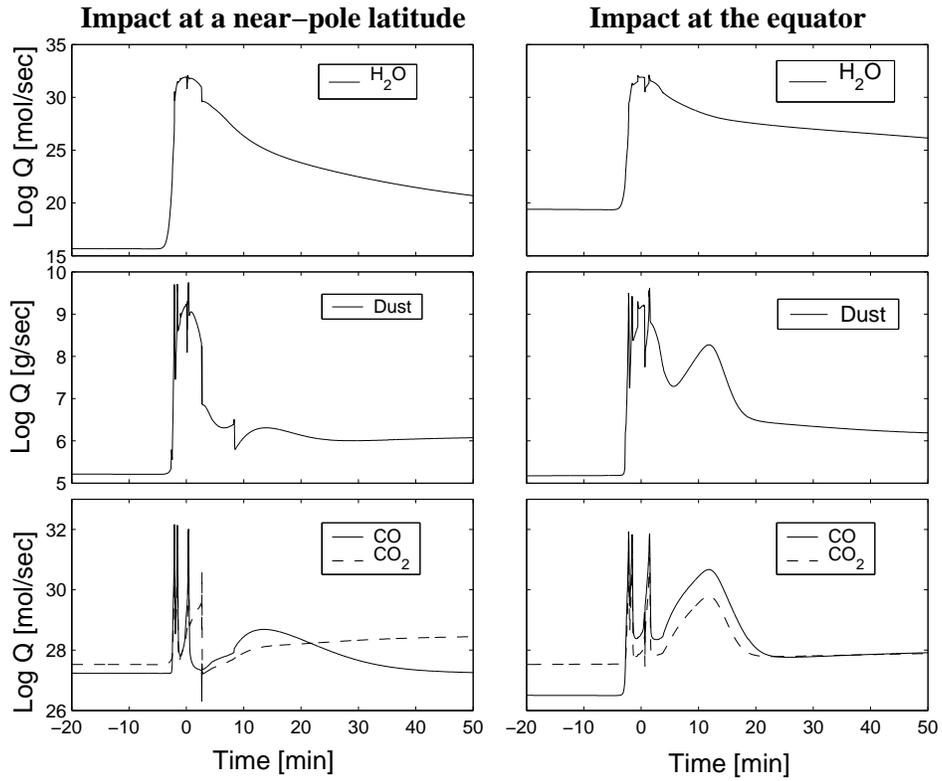}}
\caption{Comparison between the two extreme of impact latitudes: near-pole ($75.5^\circ$)
and equator ($0^\circ$). The dominant volatiles, H$_2$O, CO and CO$_2$, and the dust
component, are presented at high temporal resolution, around the peak of the impact.
Note that the general shape of production rates is similar at both impact sites.}
\label{eqpolcomp}
\end{figure}

\section{Conclusions}
\label{conclusions}

\noindent
We have followed the evolution of a rotating nucleus model in the orbit of comet 9P/Tempel 1 for
several orbits, until a quasi-steady-state was attained, and then we have simulated an impact,
similar in energy to the projectile of the {\it Deep Impact} mission, focusing on the thermal
outcome. A similar effect may be expected of a random collision with a large meteor. Provided that
such a collision will create a deep crater and expose layers a few tens of meters deep, an
outburst of gas and dust ejection is expected to result. This is due to the fact that at such depths,
ices of volatile species should be present, as well as amorphous water ice, which may
crystallize by absorbing the impact heat and release more volatiles. Our main results regarding the
outcome of the impact are as follows.
\begin{itemize}
\item{}The
increase in production rate of volatiles and dust is of several orders of magnitude, and
thus readily observable. The total dust output, for example, is equivalent to the output of about two
days at perihelion.
We should bear in mind, however, that these results provide an upper limit, since it is assumed
that the entire energy is spent in thermal effects.
\item{}An important conclusion of these
calculations is that the place on the nucleus where the impact occurs is not as significant
as one might expect. The total dust output, for example, differs by only a factor of $\sim1.3$.
\item{}Variability in the comet's activity may be detected even on the short time scale of the impact,
as the heat wave propagates into a layer of stratified composition, where amorphous water ice may
crystallize and ices of various volatile species may sublimate.
\item{}Subsequent long-term thermal evolution is also affected to some extent, but not in a way that
would be recognized by observations as an aftermath of an impact (or collision). The variation
of volatile production rates with time (or heliocentric distance)
will retain a general uneven pattern, but whether this pattern differs from that of an undisturbed comet
may be difficult to establish.
\end{itemize}
\noindent
It should be borne in mind that these predictions are based on a simplified model of a real
comet.

\bigskip
\acknowledgements
\noindent
We would like to thank Bj\"orn Davidsson for a careful reading of the manuscript
and many helpful comments.
Support for this work was provided, in part, by Israel Science Foundation grant No.942/04
and in part, through University of Maryland and University of Hawaii subcontract Z667702,
which was awarded under prime contract NASW-00004 from NASA.
\newpage

\section*{References}
\begin{description}

\item A'Hearn, M. F., Belton, M. J. S., Delamere, A., Blume, W. H. (2005) Deep Impact: A large-scale active
experiment on a cometary nucleus. Space Science Reviews, in press.

\item Belton, M. J. S., A'Hearn, M. F. (1999) Deep sub-surface
exploration of cometary nuclei. Advances in Space Research 24,
1175--1183.

\item Belton, M. J. S., Meech, K. J., A'Hearn, M. F., Groussin, O., McFadden, L., Lisse, C., Fern\'{a}ndez, Y. R.,
Pittichov\'{a}, J., Hsieh, H., Kissel, J., Klaasen, K., Lamy, P. L., Prialnik, D., Sunshine, J., Thomas, P.,
Toth, I. (2005) Deep Impact: Working properties for the target nucleus - Comet 9P/Tempel 1.
Space Science Reviews, in press.

\item Cochran, A. L., Barker, E. S., Ramseyer, T. F., Storrs, A. D.
(1992) The McDonald Observatory Faint Comet Survey - gas production
in 17 comets. Icarus 98, 151--162.

\item Cohen, M., Prialnik, D., Podolak, M. (2003) A quasi-3D model for the evolution of shape and temperature
distribution of comet nuclei - application to comet 46P/Wirtanen.
New Astronomy 8, 179--189.

\item Crifo, J. F., Rodionov, A. V. (1997) The dependence of the circumnuclear coma structure on the properties
of the nucleus. Icarus 129, 72--93.

\item Davidsson, B. J. R., Skorov, Y. V. (2002) On the light-absorbing surface layer of cometary nuclei. I.
Radiative transfer. Icarus 156, 223--248.

\item Davidsson, B. J. R., Skorov, Y. V. (2002) On the light-absorbing surface layer of cometary nuclei. II.
Thermal modeling. Icarus 159, 239--258.

\item Davidsson, B. J. R., Skorov, Y. V. (2004) Apractical tool for simulating the presence of gas comae
in thermophysical modeling of cometary nuclei. Icarus 168, 163--185.

\item Fanale, F. P., Salvail, J. R. (1984) An idealized short-period comet model - Surface insolation, H2O
flux, dust flux, and mantle evolution. Icarus 60, 476--511.

\item Farnham, T. L. (1996). PhD Thesis, University of Hawaii.

\item Fern\'{a}ndez, Y. R., Meech, K. J., Lisse, C. M., A'Hearn, M. F., Pittichov\'{a}, J., Belton, M. J. S. (2003)
The nucleus of {\it Deep Impact} target Comet 9P/Tempel 1. Icarus 164, 481--491.

\item Finson, M. L., and R. F. Probstein (1968). A theory of dust comets. Ap. J. 154, 327.

\item Fulle, M., Mikuz, H., Bosio, S. (1997) Dust environment of Comet Hyakutake 1996B2.
Astronomy and Astrophysics 324, 1197--1205.

\item Gombosi, T. I. (1994) Gaskinetic theory, Cambridge: Cambridge University Press.

\item Harker, D. E., Wooden, D. H., Woodward, C. E., Lisse, C. M. (2002) Grain properties of Comet
C/1995 O1 (Hale-Bopp). Astrophysical Journal 580, 579--597.

\item Housen, K. R. (2002) Does gravity scaling apply to impacts on
porous asteroids? Lunar and Planetary Science XXXIII, 1969.

\item Huebner, W. F., Benkhoff, J. (1999) On the relationship of chemical abundances
in the nucleus to those in the coma. Earth, Moon and Planets 77, 217--222.

\item Jockers, K. (1999) Observations of scattered light from cometary dust and their interpretation.
Earth, Moon and Planets 79, 221--245.

\item Lamy P. L., Toth, I., A'Hearn, M. F., Weaver, H. A., Weissman, P. R. (2001) Hubble Space Telescope
observations of the nucleus of 9P/Tempel 1. Icarus 154, 337--344.

\item Landolt, A. U. (1992) UBVRI photometric standards stars in the
magnitude range 11.5 $<$ V $<$ 16.0 around the celestial equator.
Astronomical Journal 104, 340--370.

\item Marsden, B. G. (1963) On the orbit of some long lost comets.
Astronomical Journal 68, 795--801.

\item Meech, K. J., A'Hearn, M. F., McFadden, L., Belton, M. J. S., Delamere, A., Kissel, J., Klassen, K.,
Yeomans, D., Melosh, J., Schultz, P., Sunshine, J., Veverka, J. (2000) Deep Impact - Exploring the
interior of a comet, in A new era in Bioastronomy, ASP Conference Series 213, G. A. Lemarchand, K. J. Meech,
eds. Astronomical Society of the pacific, 235--242.

\item Meech, K. J. (2002) The Deep Impact mission and the AAVSO. Journal of the American
Association of Variable Star Observers 31, 27--33.

\item Meech, K. J., A'Hearn, M. F., Fern\'{a}ndez, Y. R., Lisse, C. M., Weaver, H. A., Biver, N., Woodney, L. M.
(2005) The Deep Impact Earth-Based Campaign. Space Science Reviews, in press.

\item Mekler, Y., Prialnik, D., Podolak, M. (1990) Evaporation from a porous cometary nucleus.
Astrophysical Journal 356, 682--686.

\item Nolan, M. C., Asphaug, E., Melosh, H. J., Greenberg, R. (1996) Impact craters on asteroids:
does gravity or strength control their size? Icarus 124, 359--371.

\item Osip, D. J., Schleicher, D. G., Millis, R. L. (1992) Comets - Groundbased observations of
spacecraft mission candidates. Icarus 98, 115--124.

\item Podolak, M., Prialnik, D. (1996) Models of the structure and evolution of comet P/Wirtanen.
Planetary and Space Science 44, 655--664.

\item Prialnik, D. (1992) Crystallization, sublimation, and gas release in the interior of a porous comet nucleus.
Astrophysical Journal 388, 196--202.

\item Prialnik D., Benkhoff, J., Podolak, M. (2005) Modeling the Structure and Activity
of Comet Nuclei, in Comets II, M. Festou, H. U. Keller, H. A.
Weaver, eds. University of Arizona Press, 359-387.

\item Schmitt, B., Espinasse, S., Grim, R. J. A., Greenberg, J. M., Klinger,
J. (1989) Laboratory studies of cometary ice analogues. ESA SP 302
(Physics and Mechanics of Cometary Materials), 65--69.

\item Schultz, P. H., Anderson, J. L. B., Heineck, J. T. (2002) Impact crater size and evolution:
expectations for Deep Impact. Lunar and Planetary Science XXXIII, 1875.

\item Schultz, P. H., Anderson, J. L. B. (2005) Alternative cratering scenarios for the
Deep Impact collision. Lunar and Planetary Science XXXVI, 1926.

\item W\"{u}nnemann, K., Collins, G. S., Melosh, H. J. (2005) Numerical modeling of the
Deep Impact mission experiment. Lunar and Planetary Science XXXVI, 1837.

\end{description}

\end{document}